\documentclass{aa}
\usepackage{graphics}
\newcommand{\noi}{\noindent}
\def\bc{\begin{center}}
\def\ec{\end{center}}
\def\be{\begin{equation}}
\def\ee{\end{equation}}
\def\bea{\begin{eqnarray}}
\def\eea{\end{eqnarray}}
\def\ve{\varepsilon}
\def\mkm{\mu{\rm m}}
\def\ve{\varepsilon}
\def\bet{{\beta}~\rm Pictoris}
\begin{document}
\title{A new  model of composite interstellar dust grains}
\titlerunning{Composite interstellar dust grains}

\author{ N.V.~Voshchinnikov\inst{1,2},
         V.B.~Il'in\inst{1,2},
         Th.~Henning\inst{3},
         and
         D.N.~Dubkova\inst{1}
         }
\authorrunning{Voshchinnikov et al.}

\institute{
Sobolev Astronomical Institute,
St.~Petersburg University, Universitetskii prosp. 28,
           St.~Petersburg, 198504 Russia
\and
 Isaac Newton Institute of Chile, St.~Petersburg Branch
\and
Max-Planck-Institut f\"ur Astronomie, K\"onigstuhl 17, D-69117 Heidelberg, Germany
}

\offprints{N.V. Voshchinnikov, nvv@astro. \protect\linebreak
  spbu.ru}

\date{Received $<$date$>$; accepted $<$date$>$}

  \abstract{
The approach to model composite interstellar dust grains, using
the exact solution to the light scattering problem for multi-layered spheres
as suggested by Voshchinnikov \& Mathis (1999), is further developed.
 Heterogeneous scatteres are represented by particles with very large number
of shells, each including a homogeneous layer per material considered
(amorphous carbon, astronomical silicate and vacuum).
 It is demonstrated that the scattering characteristics
(cross-sections, albedo, asymmetry factor, etc.)
well converge with the increase of the number of shells (layers) and
each of the characteristics has the same limit independent of
the layer order in the shells.
 The limit obviously corresponds to composite particles
consisting of several well mixed materials. However, our results
indicate that layered particles with even a few shells (layers) have
characteristics close enough to these limits.
 The applicability of the effective medium theory (EMT)
mostly utilized earlier to approximate inhomogeneous interstellar grains
is examined on the basis of the new model.
 It is shown that the  EMT rules generally have an accuracy of
several percent in the whole range of particle sizes
provided the porosity does not exceed about 50\%.
 For larger porosity, the EMT rules give wrong results.
 Using the model, we reanalyze various basic features
of cosmic dust ---
interstellar extinction, scattered radiation, infrared radiation,
radiation pressure, etc.
 It is found that an increase of porosity typically leads to an increase
of cross-sections, albedo and the sweeping efficiency of small grains
as well as to a decrease of dust temperature and the strength of
infrared bands (the EMT fails to produce these effects).
 As an example of the potential of the model, it is applied to reproduce
the extinction curves in the directions to $\zeta$ Oph and $\sigma$ Sco
using subsolar cosmic abundances.
 We also conclude
that metallic iron even in negligible amount ($\la 1$\,\% by the volume fraction)
is unlikely to form a layer on or inside a grain
because of peculiar absorption of radiation by such particles.
      }
\maketitle

\section{Introduction}
Scientists
felt  the  necessity to treat the scattering by
composite and inhomogeneous
particles and media
(i.e., consisting of several components)
essentially earlier than
the existence of interstellar dust has been generally established.
It was started by
Garnett~(\cite{gar04}) who found
the averaged or effective dielectric functions
of a  medium  assuming that one material
was a matrix (host material) in which the other material was embedded
(so called   Maxwell--Garnett mixing rule of Effective Medium Theory; EMT).
When the roles of the inclusion and the host material are reversed,
the inverse Garnett rule is obtained.
Later, Bruggeman~(\cite{brug35}) deduced
another rule which
was symmetric with respect to the interchange of materials.
These classical mixing rules are the most popular ones till now.

Many scientific and applied problems require
exact calculations of light scattering by inhomogeneous particles.
This became first possible in the beginning of the 1950s when
the Mie solution for homogeneous spheres was
generalized to core-mantle spherical particles in three independent
papers (Aden \& Kerker \cite{ak51},
Shifrin \cite{s52}, G\"uttler \cite{gut52}).
G\"uttler's solution was used by
Wickramasinghe~(\cite{w63}) who first calculated the extinction of
graphite core-ice mantle  grains.
Such particles should be formed in interstellar clouds where the bare
particles ejected from  stellar atmospheres
may accumulate mantles from volatile elements.

An unsuccessful first attempt to detect the 3.1\,$\mkm$ feature
of solid H$_2$O together with pioneering experiments
on the ultraviolet (UV) photoprocessing of mixtures of
volatile molecules stimulated Greenberg~(\cite{g84})
to suggest the idea of the formation
of complex organic molecules on silicate cores. In dense clouds,
an additional layer of photolysed volatile ices appears due to accretion,
so the particles become three-layered.
Further, the grains around protostellar objects are covered by a
layer of unphotolysed accumulated atoms and molecules, then the particles
become four-layered.
This core-mantle model was modified several times
by Greenberg and his co-workers
(Hong \&  Greenberg \cite{hg80},
Greenberg \&  Li \cite{gl96a},
Li \&  Greenberg \cite{lg97},
see also a brief history of development of dust models
in Li \&  Greenberg \cite{lg02}).
Another model of inhomogeneous interstellar grains was
proposed by Duley et al.~(\cite{djw89}) who considered
silicate grains coated by thin layers of hydrogenated amorphous carbon
(HAC) and amorphous carbon.
Although  the possible origin of such grains remains unclear,
they were applied to the explanation of interstellar extinction,
extended red emission, unidentified infrared (IR) bands, etc.
Jones et al.~(\cite{jdw90}) showed that the
evolution of HAC mantle in the interstellar conditions
could result in the appearance of
a layer of polymeric HAC overlaying the graphitic HAC which
in its turn covered the silicate core.
However,  the optical properties of three and more layered particles
were never calculated exactly: usually some EMT or, in the best case,
a core-mantle grain model was used.

Core-mantle (or even multi-layered) grains are possibly produced
in  the atmospheres of late-type stars.
McCabe~(\cite{mc82})  suggested that
because of the  inverse greenhouse effect  silicon carbide core-graphite
mantle grains could be produced in carbon stars.
This idea was further developed by  Kozasa et al.~(\cite{ko96}) who demonstrated that
nucleation of SiC grains always preceded that of carbon grains
if the difference between the temperature of gas and small clusters
was taken into account.   The application of
this model to interpretation of the spectral energy distribution
of carbon stars show that the core-mantle grains better fit the data for
dust envelopes of evolved stars than the mixtures of homogeneous
grains which are  able to reproduce the data for optically thin envelopes
(Lorenz-Martins et al. \cite{lm01}).

Mathis \&  Whiffen~(\cite{mw89})
introduced the model of composite grains which were
very porous (the volume fraction of vacuum $\sim 80$\,\%)
aggregates of small amorphous carbon, silicate and
iron particles. The optical properties of such particles
were calculated with the Mie theory and EMT.
Mathis~(\cite{m96}) updated the composite grain model
taking into account new subsolar abundances of heavy elements.
The new model consisted of three components where the visual/near-IR
extinction was explained by aggregates with $\sim 45$\,\% vacuum
in volume.

Now light scattering computations for inhomogeneous (composite) particles with
layers or inclusions from different materials  or aggregate particles
can be made using the discrete dipole approximation (DDA),
the T-matrix method (TMM) or a simpler theory like the  Mie theory
for $n$-layered spheres (see Voshchinnikov \cite{v02} for discussion).
However, calculations with the DDA  are very time-consuming and at the present
can be used rather for illustrative than  mass calculations (e.g.,
Wolff  et al. \cite{wcms-l94}, Vaidya  et al. \cite{vgdc01}).
The idea of composite particles as multi-layered spheres
(Voshchinnikov \&  Mathis \cite{vm99},
see also Iat\`\i \,  et al. \cite{iatietal01}) looks
a bit artificial but  attractive from the point of
view of numerical realization.
Such a model permits to include an arbitrary
fraction of any material and the computations  require rather moderate
resources.

In this paper, we develop the model of  composite interstellar
grains based on  {\it exact} calculations for multi-layered spheres.
The general description of the model is given in Sect.~\ref{model}.
We calculate  different efficiency factors, albedo, etc.
and analyze how they depend on the order and number of layers
(Sect.~\ref{eff}) and the fraction of vacuum (Sect.~\ref{vac}).
The possibility to describe the optical properties of
multi-layered spheres using Mie theory
with different EMT rules is considered in Sect.~\ref{emt}.
The wavelength dependence of extinction by multi-layered particles
is discussed in Sect.~\ref{ext1}.
The analysis of the radiation pressure on composite grains
is presented in Sect.~\ref{pr}.
The behaviour of albedo and asymmetry parameter as well as
the intensity and polarization of the scattered radiation
are studied in Sect.~\ref{sca}.
The next section includes the consideration of grain temperature
(Sect.~\ref{temp}), profiles of IR bands (Sect.~\ref{ir_b}),
and grain opacities at $\lambda = 1$\,~mm (Sect.~\ref{opa}).
Section~\ref{abun} contains the application of the new model to
the calculations of the extinction  curves in the direction
of two stars taking into account the subsolar cosmic abundances.
We also analyze the possibility of pure iron to be a component of
a multi-layered particle (Sect.~\ref{fe}).
Concluding remarks are presented in Sect.~\ref{concl}.

Further development of the model will involve the consideration of
non-spherical  multi-layered grains based on the
available light scattering methods (see Farafonov et al. \cite{fip02}
for a review).

\section{General description of the model}\label{model}

\subsection{Basics}

We construct  composite grains as particles consisting of many concentric
spherical layers of various materials, each with a specified
volume fraction. Vacuum can be one of the materials, so a
composite particle may have a central cavity or voids in the form
of concentric layers.
From the point of view of dust formation and growth,
the presence of vacuum at the particle centre
or in several voids distributed inside porous aggregate grain is rather natural
while a particle with  concentric spherical vacuum layers looks
artificial. However, in the case of multi-layered composite spheres we
can include at {\it any} position inside
a particle {\it any} fraction of a material
(from extremely small to very large) and produce the {\it exact}
calculations. The latter is of particular importance for the consideration
of very porous grains as suggested, for example,
for comets (Greenberg \& Hage~\cite{grha91})
or the  disc of $\bet$ (Li \& Greenberg~\cite{li:gre98}).

\begin{figure}[htb]
\bc
\resizebox{8.8cm}{!}{\includegraphics{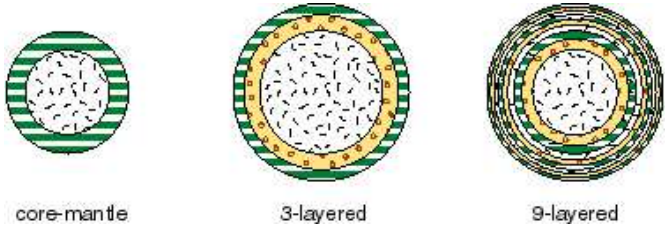}}
\caption{
The models of composite particles containing the same
amount of two materials (carbon and silicate).
The multi-layered spheres consist of an equal volume fraction (33.33\%)
of carbon, silicate and vacuum separated in equivolume spherical layers.
The core-mantle particle has the same mass of carbon and
silicate but is free of vacuum.
As a result, its volume is less by 1/3 and the outer radius is
by $\sqrt[3]{1/3} \approx 0.69$ less.
}\label{f00}\ec\end{figure}

The schematic representation of multi-layered spherical grains
is given in Fig.~\ref{f00}.
They are composed of a specified number of
concentric spherical homogeneous layers.
The amount of  material is determined by its volume  fraction
$V_i$ ($\Sigma_i V_i /V_{\rm total} = 1)$.
The amount of vacuum (or the particle porosity ${\cal P}$, $0 \leq {\cal P} < 1$)
can be introduced as
\be
{\cal P} = V_{\rm vac} /V_{\rm total}
= 1 - V_{\rm solid} /V_{\rm total}. \label{por}
\ee
The  order of the layers and their total number can be specified
separately.
Following Voshchinnikov \&  Mathis~(\cite{vm99}, hereafter VM),
we assume further that the concentric spherical layers of two or more
different materials form a ``shell". The whole grain
consists of a specified number of concentric shells, e.g.
the simplest composite
particle contains one shell of two materials (core-mantle or coated
grain). As materials, we choose carbon and silicate
which were used in many cosmic dust models
(Mathis et al. \cite{mrn}, Draine \& Lee \cite{dl84}).
The illustrations of several composite grains are presented in Fig.~\ref{f00}.
The core-mantle particle does not contain vacuum, but its mass
is the same as that of the other two particles and, therefore,
its outer radius is smaller.
Note also that since the volume fractions are specified, the
innermost layer is relatively thicker than the others.

The formal solution to the light scattering problem
for multi-layered spheres can be
easily written in  matrix form (see, e.g., Kerker \cite{k69}).
However, for practical reasons, it is better to use
the recursive algorithm developed by Wu \& Wang~(\cite{ww91}) and
Johnson~(\cite{j96}). In order to make  calculations for highly
absorbing particles of large size, several modifications
were suggested by Wu et al.~(\cite{wuetal97}) and
Gurvich et al.~(\cite{gsk01}).

In our calculations presented below, composite particles of
several materials are considered. The refractive indices for them
were taken from the Jena--Petersburg Database of Optical Constants (JPDOC)
which was described by Henning et al.~(\cite{heal99}) and
J\"ager et al.~(\cite{jetal02}).

\subsection{Dependence on the order and number of layers}\label{eff}

\begin{figure}\bc
\resizebox{\hsize}{!}{\includegraphics{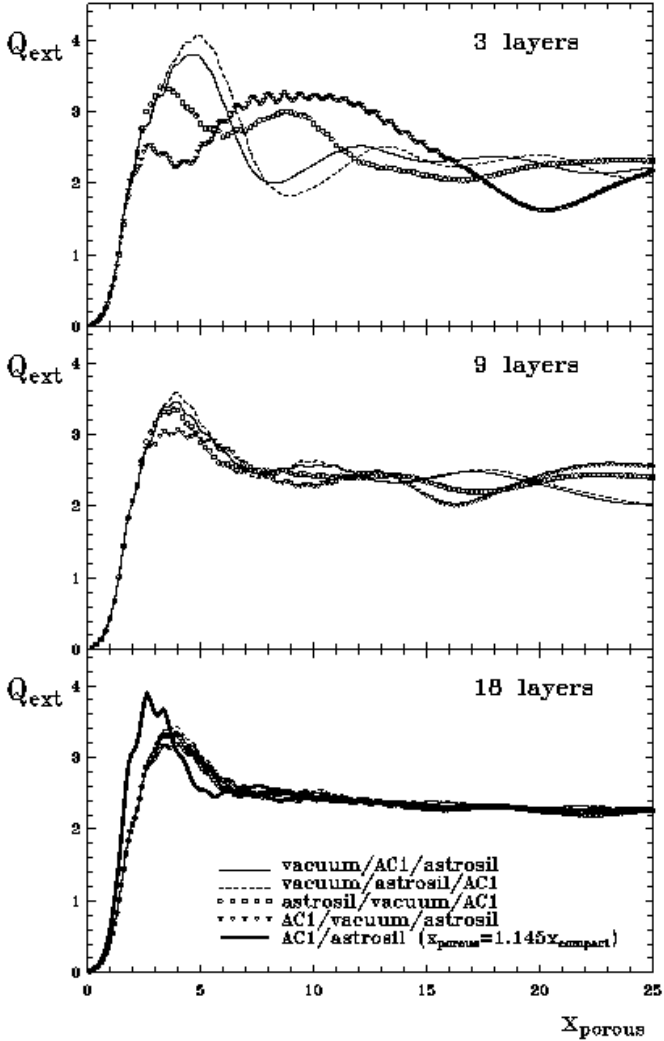}}
\caption{Size dependence of the extinction efficiency factors
for multi-layered spherical particles.
The size parameter $x_{\rm porous}$ is calculated according to Eq.~(\ref{xpor}).
Each particle contains an equal fraction (33.33\%)
of amorphous carbon (AC1), astrosil and  vacuum separated in equivolume layers.
The porosity of the particles with 3, 9 and 18 layers is the same
(${\cal P}=1/3$).
The cyclic order of the layers is indicated (starting from the core).
The effect of the increase of the number of layers is illustrated.
The thick line at the lowest panel corresponds to compact spheres
consisting of AC1 and astrosil. For a given value of the size parameter,
the compact and porous particles have the same mass. In order to
reach that, the $x$ scale for compact particles was stretched
by a factor $\sqrt[3]{3/2} \approx 1.145$.
}\label{f1}
\ec
\end{figure}

The optical properties of core-mantle spheres have been studied rather well
and seem to show no significant peculiarities (Prishivalko  et al.~\cite{pbk84}).
In contrast, already three-layered spheres
can produce  anomalous extinction of light. This is illustrated in
Fig.~\ref{f1} where the extinction efficiency factors $Q_{\rm ext}$
are plotted for spheres consisting of 3, 9 and 18  equivolume layers.
The layers are composed of
amorphous carbon (AC1), astronomical silicate (astrosil) and vacuum,
the volume fraction of each constituent is 1/3.
The optical constants for AC1 ($m=1.98+0.23i$) and astrosil ($m=1.68+0.03i$)
correspond to the wavelength $\lambda=0.55\,\mkm$ and
were taken from the papers of Rouleau \& Martin~(\cite{rm91}) and
Laor \&  Draine~(\cite{laordr93}), respectively.

\begin{figure}\bc
\resizebox{\hsize}{!}{\includegraphics{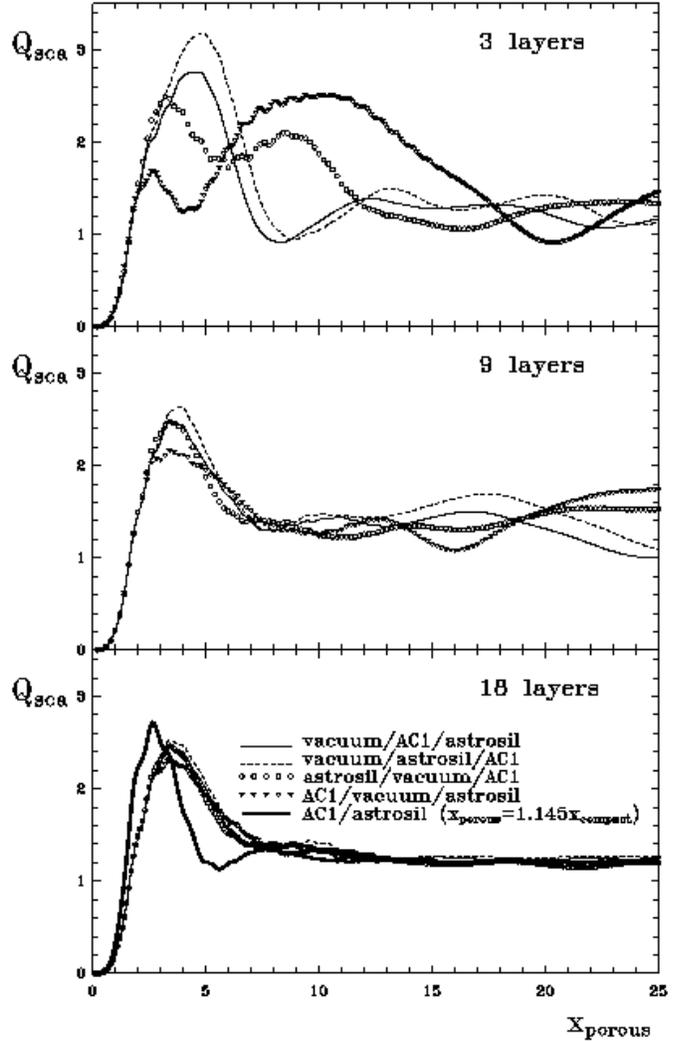}}
\caption{The same as in Fig.~\protect\ref{f1} but now for scattering efficiency factors.}
\label{f2} \ec \end{figure}

\begin{figure}\bc
\resizebox{\hsize}{!}{\includegraphics{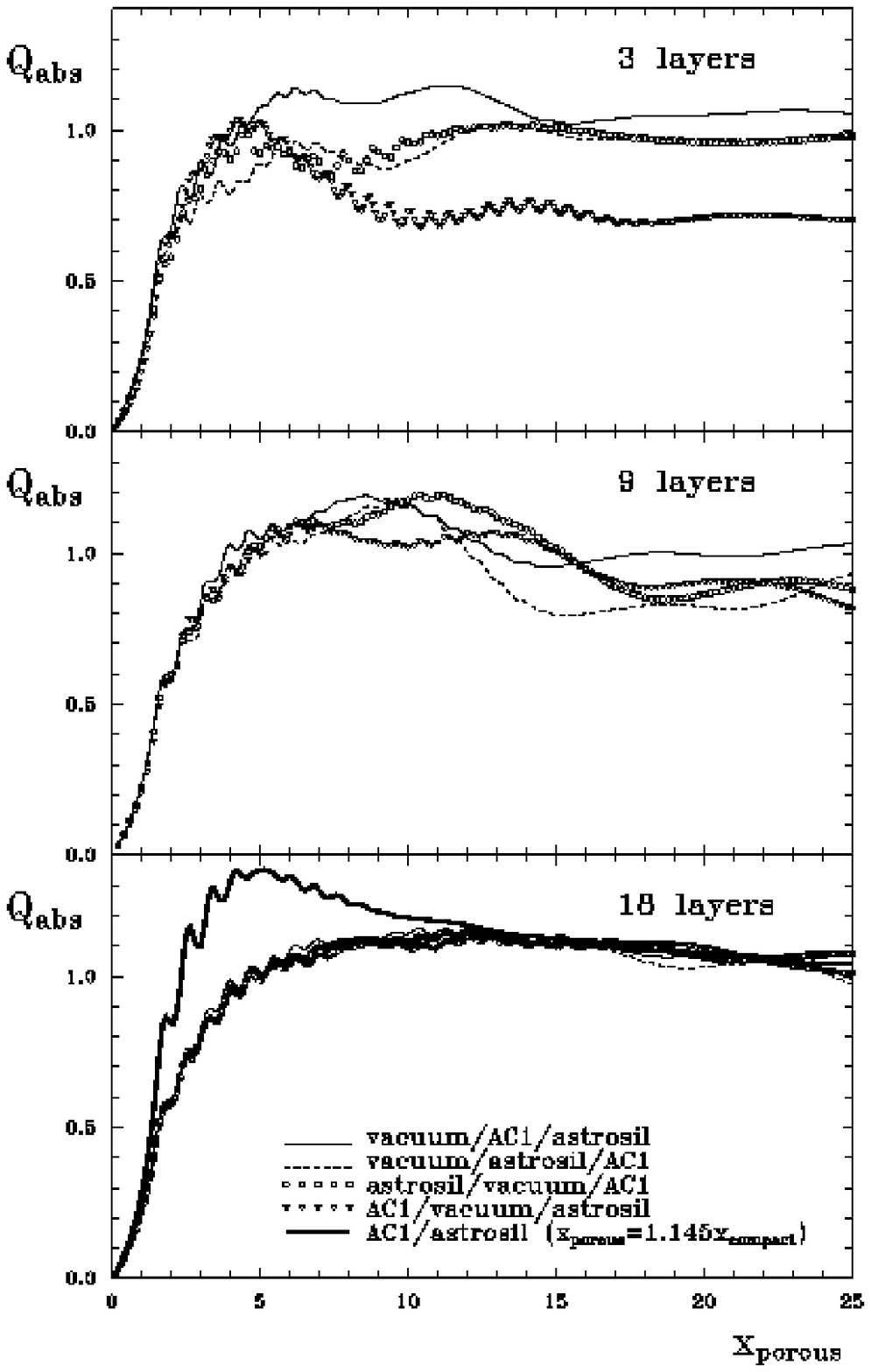}}
\caption[]
{The same as in Fig.~\ref{f1} but now for absorption efficiency factors.}
\label{f3}\ec \end{figure}

\begin{figure}\bc
\resizebox{\hsize}{!}{\includegraphics{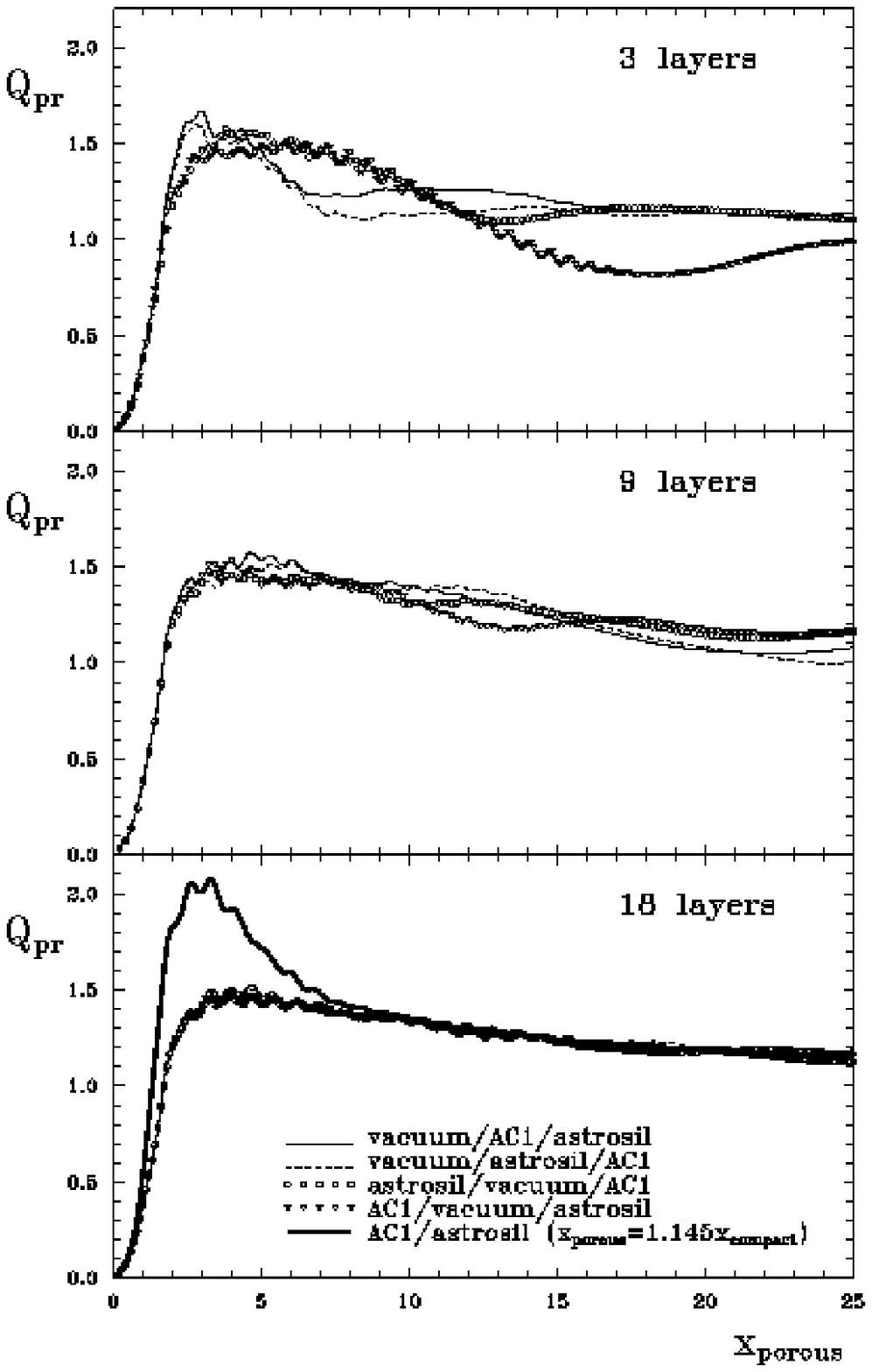}}
\caption[]
{The same as in Fig.~\ref{f1} but now for radiation
pressure efficiency factors.}
\label{f4}\ec \end{figure}

\begin{figure}\bc
\resizebox{\hsize}{!}{\includegraphics{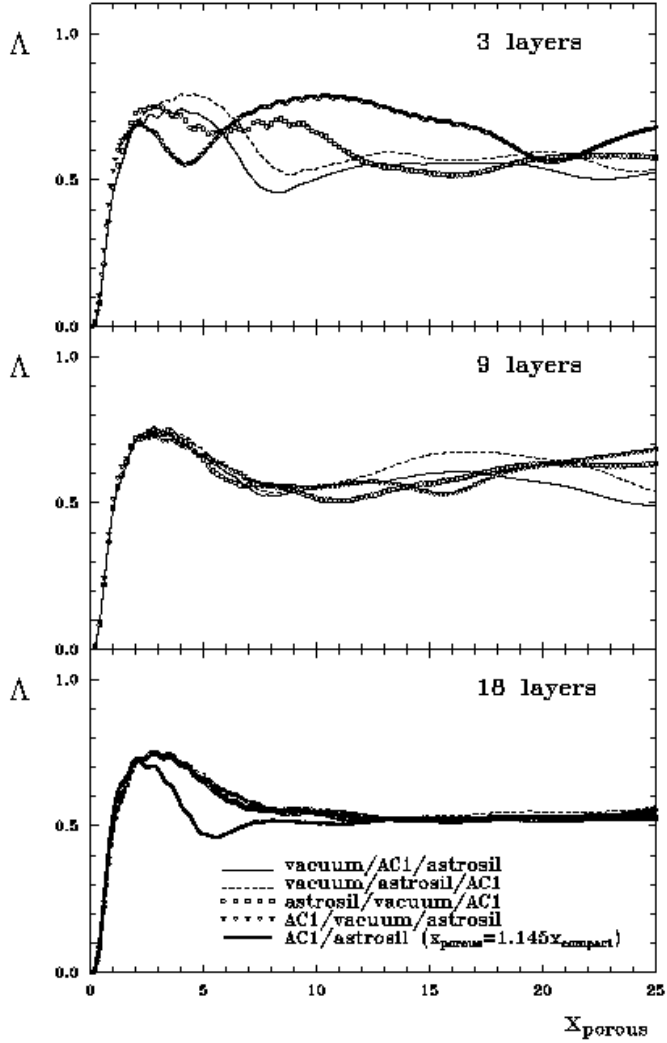}}
\caption[]
{The same as in Fig.~\ref{f1} but now for albedo.}
\label{f5}\ec \end{figure}

\begin{figure}\bc
\resizebox{\hsize}{!}{\includegraphics{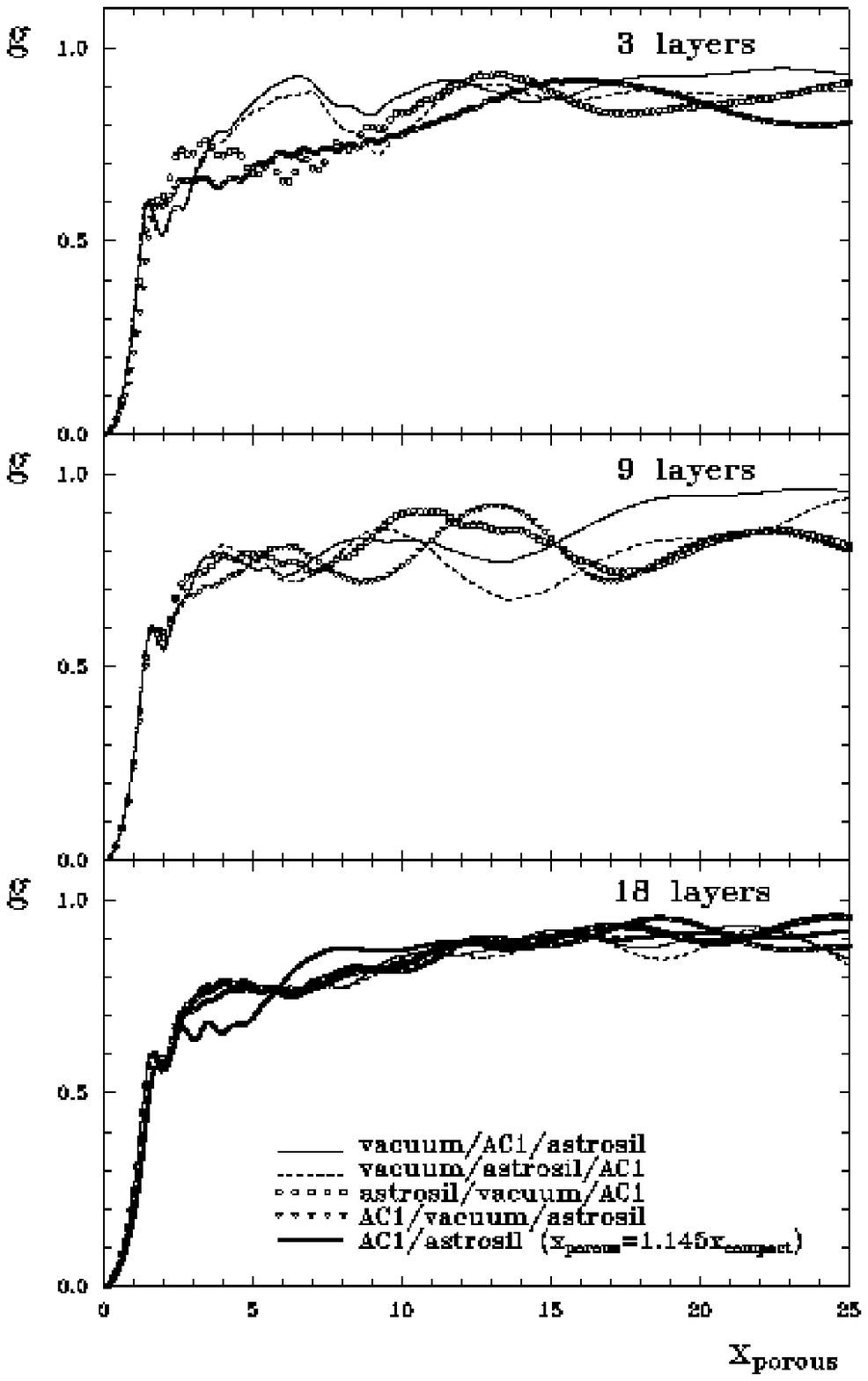}}
\caption[]
{The same as in Fig.~\ref{f1} but now for asymmetry parameter.}
\label{f6}\ec  \end{figure}

The order of the materials strongly affects
the behaviour of extinction in the case of three-layered particles
(the upper panel of Fig.~\ref{f1}). First of all, the position of vacuum
(the core or the middle layer) is important.
The  curve for the case of particles
with a carbon core and an outermost astrosil layer is the most peculiar curve.
Here, a very rare situation is observed:
The first maximum is damped,
but there is a very broad second maximum which is the highest
among the different cases.
As follows from the upper panels of Figs.~\ref{f2} and \ref{f3},
the scattering efficiency depends stronger on
the order of layers than the absorption efficiency.
However, all the peculiarities
disappear if the number of layers increases: The difference
between the curves becomes rather small for particles with 9 layers (3 shells)  and
is hardly present for particles with 18 layers
(6 shells; see the middle and lower panels of Figs.~\ref{f1}~--~\ref{f6}).
This fact, noted by VM already, allows one {\it to use multi-layered
particles as a new approximate model of composite grains}.
Such a model is the {\it only} possible  way  to treat
heterogeneous particles exactly  when several materials are well mixed.
Here the particle has to be divided in many shells ($\ga 3 - 5$).
Then we obtain a ``composite'' particle with ``average'' optical properties
where the real order of materials in each shell can be ignored.
For such particles, different efficiency factors as well as
the particle  albedo $\Lambda$ and
the asymmetry parameter $g$ (or $\langle \cos \Theta \rangle$)
depend on the volume fraction of materials only.
Below we consider a particle consisting of 18 layers as
a representative of a heterogeneous interstellar grain with well mixed constituents.

Solid thick lines at the lowest panels of Figs.~\ref{f1}~--~\ref{f6}
show the size dependence of the optical characteristics for compact
(${\cal P} = 0$) spheres consisting of the same amount of solid materials
$V_{\rm solid} /V_{\rm total}$. To compare the
optical properties of porous and compact particles,
one needs to normalize the size parameter
(or radius) of the compact (or porous) particle using the relation
\be
x_{\rm porous} = \frac{x_{\rm compact}}{(1-{\cal P})^{1/3}}
= \frac{x_{\rm compact}}{(V_{\rm solid} /V_{\rm total})^{1/3}}. \label{xpor}
\ee
In the case of the particles presented in Figs.~\ref{f1}~--~\ref{f6},
this leads to stretching of the $x$ scale for compact particles
by a factor of $\sqrt[3]{3/2} \approx 1.145$.
It can be seen that the presence of vacuum inside a composite particle
reduces the peak of the absorption efficiency (Fig.~\ref{f3})
and shifts  that of the scattering efficiency (Fig.~\ref{f2}).
Correspondingly,  these two effects explain
the behaviour of the curves for the extinction and radiation pressure factors
(Figs.~\ref{f1} and \ref{f4}).
A medium porosity  influences the albedo and the
asymmetry parameter only in a restricted range of the size parameters
$x_{\rm porous} \approx 3 - 10$ (see Figs.~\ref{f5} and \ref{f6}),
but the properties of very porous  and compact particles
of all sizes differ considerably (see discussion in the next Section).

\subsection{Varying the fraction of vacuum}\label{vac}

The fraction of vacuum in interstellar dust grains can be large.
Very porous particles are often used to model cometary grains.
For example, Greenberg \& Hage~(\cite{grha91}) indicate
that the porosity of dust aggregates in comets can be in the range
$0.93 < {\cal P} < 0.98$. Their conclusions are based on the
model of porous grains developed by Hage \& Greenberg~(\cite{hagr90})
who used for light scattering calculations
the volume integral equation formulation method
(a modification of DDA). The verification of this method
was made in the case of small compact spheres but later
the method was generalized to large and very porous particles.
A qualitative agreement of the method
with the Mie--Garnett calculations was found: in both cases the
absorption efficiency cross-sections
$C_{\rm abs}$  increase while the albedo $\Lambda$ decreases when
the porosity grows. Although the validity of these conclusions
for particles beyond the Rayleigh domain remains unclear,
the results of Hage \& Greenberg~(\cite{hagr90}) are  frequently used for
estimates of grain properties in comets
(see, e.g., Mason et al.~\cite{masetal01}).

The role of porosity in dust optics was also analyzed by
Kr\"ugel \& Siebenmorgen~(\cite{ks94}) who calculated the normalized
absorption cross-sections
\bea
C^{\rm (n)} = \frac{C({\rm porous \, grain})}
{C({\rm  compact \, grain \, of \, same \, mass})} = \nonumber \\
\,\,\,\,\,\,\, (1-{\cal P})^{-2/3}\,  \frac{Q({\rm porous \, grain})}
{Q({\rm  compact \, grain \, of \, same \, mass})}. \label{cn}
\eea
Effective optical constants of porous particles were calculated,
in particular, using the Bruggeman mixing rule.
The Mie theory was applied to get  $Q_{\rm abs}$.
Kr\"ugel \& Siebenmorgen found that the cross-sections
$C^{\rm (n)}_{\rm abs}$ increased until ${\cal P} \la 0.6$ and then decreased.

\begin{figure}\bc
\resizebox{\hsize}{!}{\includegraphics{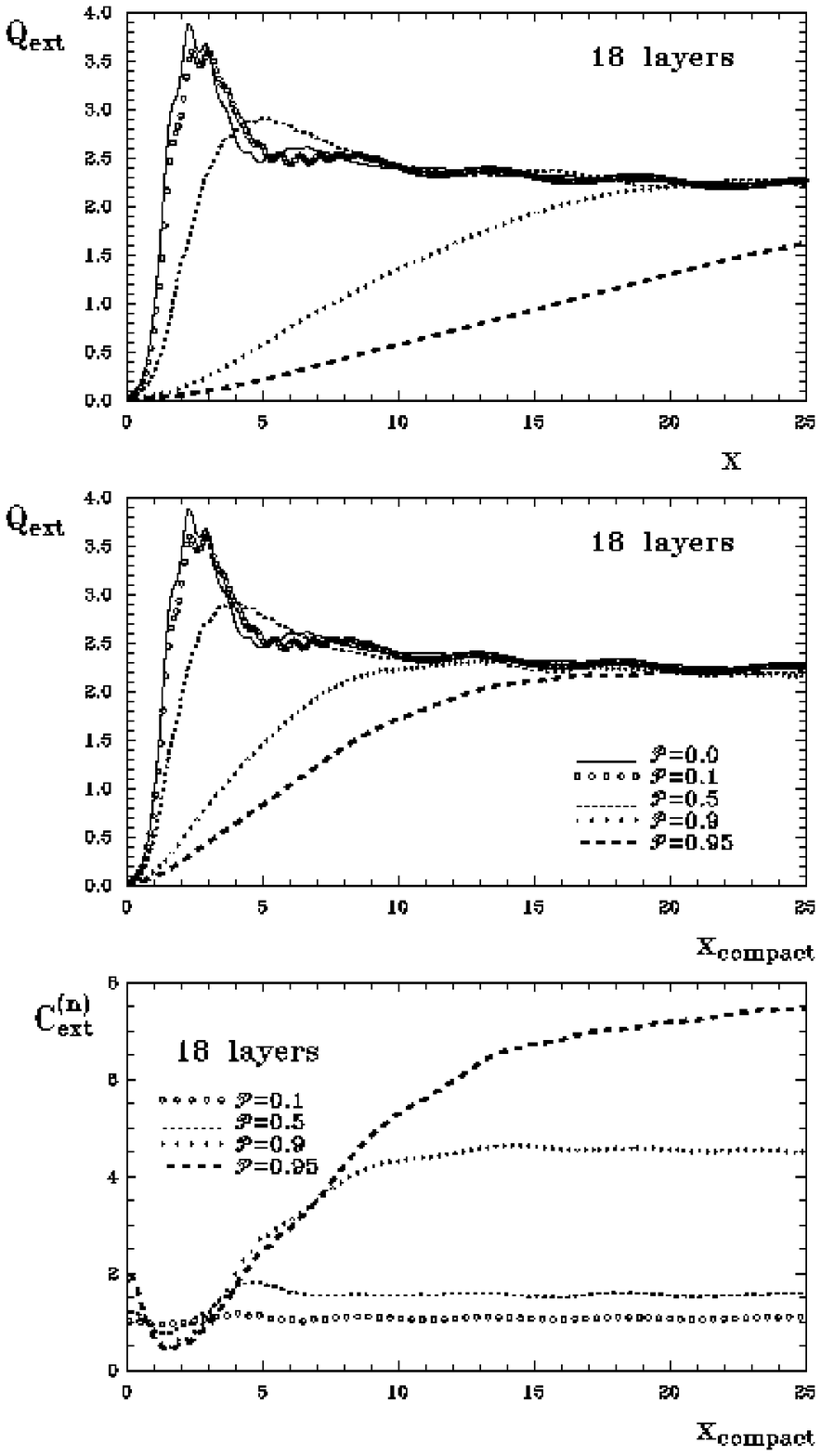}}
\caption
{Size dependence of the extinction efficiency factors
for multi-layered spherical particles.
Each particle consists of 18 layers (6 shells).
Each shell contains the same volume fractions
of AC1 and astrosil and a varied fraction of vacuum.
The cyclic order of the layers is AC1/vacuum/astrosil.
The effect of the increase of the particle porosity is illustrated.
The factors $Q_{\rm ext}$ are plotted in dependence on
the size parameter corresponding the outer particle radius
(upper panel) and the size parameter corresponding to the volume
fraction of solid materials (see Eq.~(\ref{xpor}); middle panel).
The lower panel shows the normalized extinction cross-sections
calculated according to Eq.~(\ref{cn}).}
\label{fvac}\ec \end{figure}

\begin{figure*}\bc
\resizebox{\hsize}{!}{\includegraphics{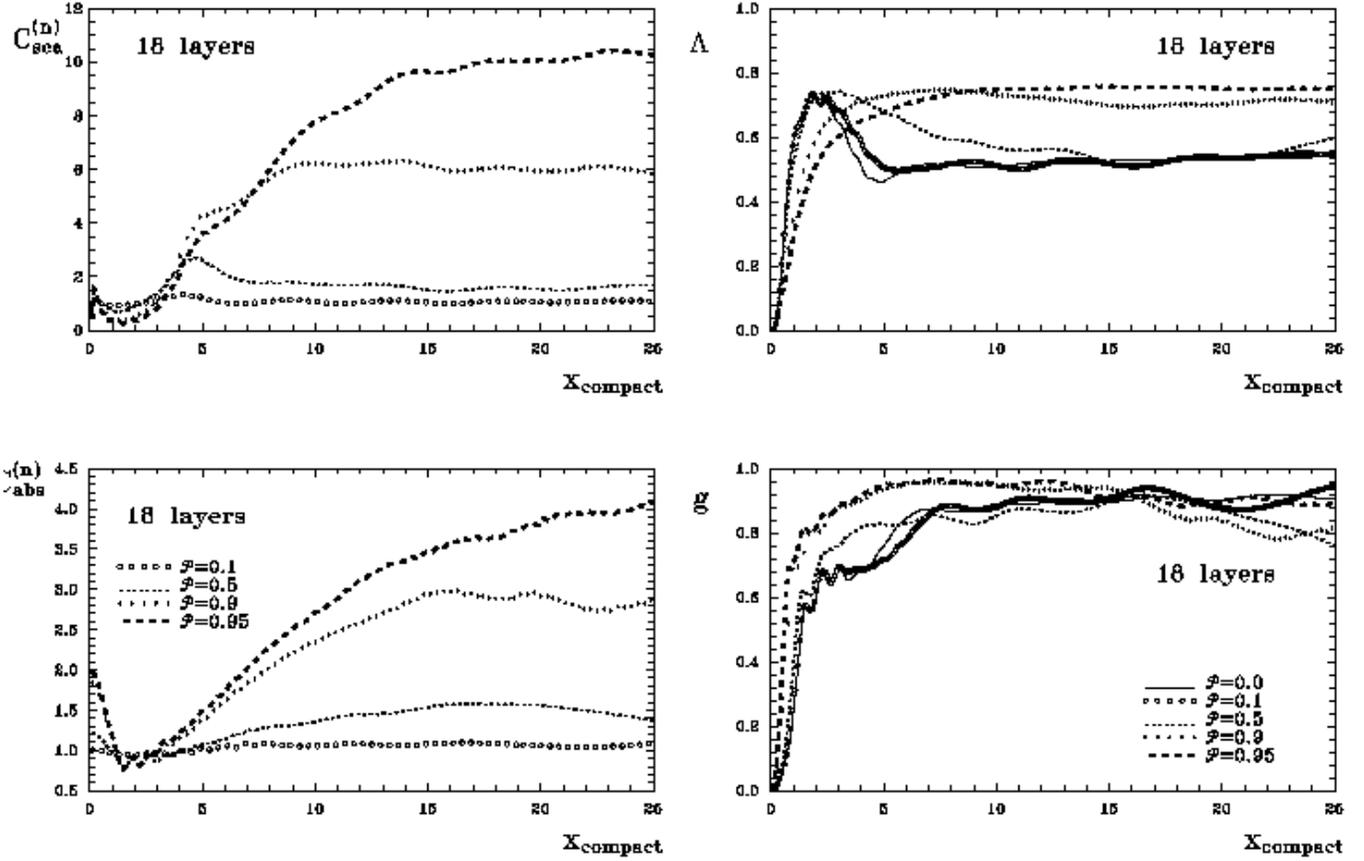}}
\caption
{Size dependence of the scattering and absorption normalized cross-sections,
albedo and asymmetry parameter for multi-layered porous spheres.
The parameters of particles are the same as in Fig.~\ref{fvac}.
}\label{fvac2}\ec \end{figure*}

Figure~\ref{fvac} shows the extinction efficiencies and normalized
cross-sections for multi-layered spheres of increasing porosity.
The results are plotted in two scales: related to
the real outer particle size parameter $x$ and the size parameter
calculated from the volume fraction of solid materials $x_{\rm compact}$
(Eq.~(\ref{cn})).
It can be seen that the growth of porosity leads to
the disappearance of the first maximum: the curves
immediately approach  the limiting value $Q_{\rm ext}=2$.
It is important that the extinction factors generally decrease when ${\cal P}$
increases\footnote{A similar conclusion can be made
from the DDA calculations of Wolff  et al.~(\cite{wcms-l94})
who considered  particles with ${\cal P}$ up to 0.8.}
and the values of $C^{\rm (n)}_{\rm ext}$ almost
always are greater than unity, i.e.  porosity increases  extinction.
An opposite case is observed only in a restricted range
of the size parameters $x_{\rm compact} \approx 1 - 3$.

As follows from Fig.~\ref{fvac2}, such a behaviour
of the extinction cross-sections is combined with the similar
effects in scattering and  absorption cross-sections.
At the same time, the scattering efficiencies slightly grow
and the absorption efficiencies significantly decrease 
for very porous grains.
Note also that both $\Lambda$ (beginning with
size parameter $x_{\rm compact} \ga 2 - 3$)
and $g$ (for particles of all sizes) increase with the particle porosity.
The  behaviour of $C^{\rm (n)}_{\rm abs}$ and $\Lambda$
found by us is more complicated than predicted by Hage \& Greenberg~(\cite{hagr90}).
Namely, the growth of porosity leads to an increase of $C^{\rm (n)}$
and an decrease of  $\Lambda$ for very small grains,
and to an increase of both quantities for large grains.
There exists also a small interval of particles of intermediate sizes where
both $C^{\rm (n)}$ and $\Lambda$ decrease.

\section{On the possiblity to describe the optical properties
of multi-layered particles with Effective Medium Theory}\label{emt}

\begin{figure*}[htb]\bc
\resizebox{\hsize}{!}{\includegraphics{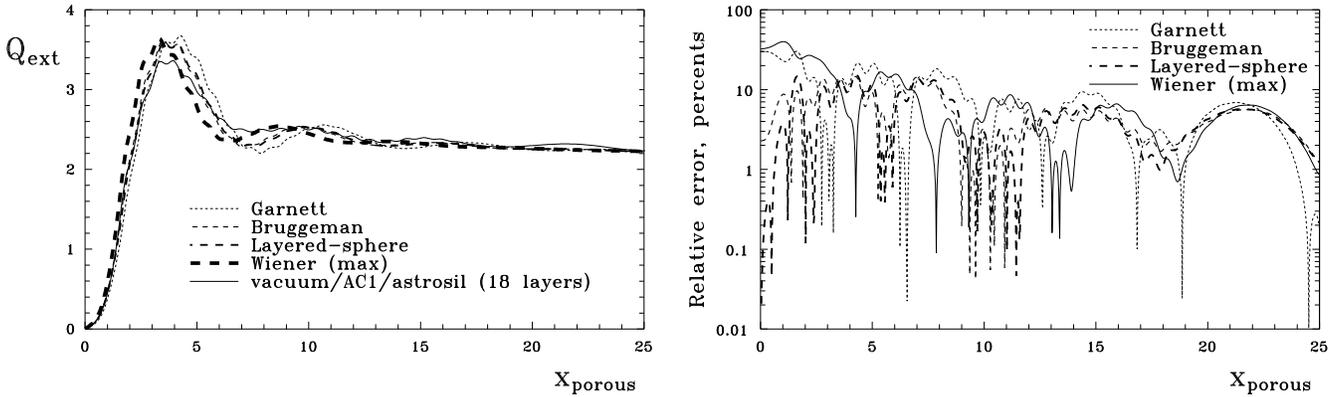}}
\caption{
   Size dependence of the efficiency factors (left panel)
   and their relative errors (right panel)
   calculated with the exact theory for multi-layered spheres
   and with the Mie theory using four different EMT rules.
   Multi-layered particles contain an equal fraction  of
   amorphous carbon (AC1), astrosil and vacuum.
   The cyclic order of the 18 layers is indicated.
}\label{emt1}\ec\end{figure*}

The EMT is an approach to treat
inhomogeneous scatterers by
homogeneous particles having an average (effective)  refractive index.
The EMT is well described in the recent reviews of
Sihvola~(\cite{sih99}), Ch\'ylek et al.~(\cite{cval00})
and  papers of Spanier \& Herman~(\cite{sh00}) and
Kolokolova \& Gus\-taf\-son~(\cite{kolgust01}).
 There are many EMT rules, but besides a few
ones they are rather similar in principle.
 Here we give  formulas of the most often used EMT rules
for $n$-component mixtures:
the Garnett~(\cite{gar04}) and Bruggeman~(\cite{brug35}) rules.
In the first case, the mixing rule averages the
dielectric permittivities of inclusion materials $\ve_{i}$
and a ``matrix'' (host) material
$\ve_{\rm m}$$^($\footnote{The dielectric permittivity
is related to the refractive index as $\ve=m^2$.}$^)$
\be
{\ve}_{\rm eff} = \ve_{\rm m} \left( 1 +
\frac{ 3 \sum_{i} f_{i}
\displaystyle\frac{\ve_{i} - \ve_{\rm m}}{\ve_{i} + 2 \ve_{\rm m}}}
{ 1 - \sum_{i} f_{i}
\displaystyle\frac{\ve_{i} - \ve_{\rm m}}{\ve_{i} + 2 \ve_{\rm m}}}
\right),
\ee
where $f_{i}=V_i /V_{\rm total}$ is the volume fraction
of the $i$th material and $\ve_{\rm eff}$ is the effective permittivity.
The expression for the Bruggeman~(\cite{brug35}) rule is
\be
\sum_i f_{i} \frac{\ve_{i} - \ve_{\rm eff}}{\ve_{i} + 2 \ve_{\rm eff}} = 0.
\ee
As an example of a more sophisticated rule,
we use the ``layered-sphere EMT'' introduced by VM.
In this case, the effective optical constants $\ve_{\rm eff}$
are defined as follows (see also Farafonov \cite{f00}):
\be
\ve_{\rm eff} = \frac{1+2\alpha/V}{1-\alpha/V} =
   \frac{{\cal A}_2}{{\cal A}_1}, \label{ls1}
\ee
where $\alpha$ is the complex electric polarizability and
the coefficients ${\cal A}_1$ and ${\cal A}_2$ are obtained
as a result of multiplication of matrices depending on the optical
constants of layers and the volume fractions
\bea
\left( \begin{array}{c}
{\cal A}_1 \\ {\cal A}_2
\end{array} \right) & = &
\left( \begin{array}{cc}
1 & {1}/{3} \\
\ve_n & -{2}/({3}\ve_n)
\end{array} \right)  \nonumber \\
& \times & \prod_{i=2}^{n-1}
\left( \begin{array}{cc}
{1}/{3} \left( \frac{\ve_{i}}{\ve_{i+1}} + 2 \right) &
-{2}/({9 f_{i}}) \left( \frac{\ve_{i}}{\ve_{i+1}} - 1 \right) \\
-f_{i} \left( \frac{\ve_{i}}{\ve_{i+1}} - 1 \right) &
{1}/{3} \left( 2 \frac{\ve_{i}}{\ve_{i+1}} + 1 \right)
\end{array} \right)
\nonumber \\
& \times &
\left( \begin{array}{c}
{1}/{3} \left( \frac{\ve_{1}}{\ve_{2}} + 2 \right) \\
- f_{i} \left( \frac{\ve_{1}}{\ve_{2}} - 1 \right),
\end{array} \right). \label{ls2}
\eea

The absolute bounds to $\ve_{\rm eff}$ were given by Wiener (\cite{wiener10})
\be
\ve_{\rm eff, \, max} = \sum_i f_i \ve_i,
\label{w1}
\ee
and
\be
{\ve_{\rm eff, \, min}} = \left( \sum_i \frac{f_i}{\ve_i} \right)^{-1}.
\label{w2}
\ee
For any composition and structure,
$\ve_{\rm eff}$ has not to lie beyond these limits
as long as the microstructural dimensions remain small compared with
the radiation wavelength.
Note that in the composite grain model of Mathis \&  Whiffen~(\cite{mw89})
the effective refractive index was calculated from Eq.~(\ref{w1}),
i.e. the maximum  of the permissible refractive indices was taken.

The general condition of EMT applicability is
that the size of ``inclusions'' (in the EMT the particle inhomogeneity
is considered in the form of uniformly distributed inclusions)
is small in comparison with
the wavelength of incident radiation (Ch\'ylek et al. \cite{cval00}).
The real range of applicability of different rules was shown
to be nearly the same (see, e.g., Table~4 in Voshchinnikov \cite{v02}).

\begin{figure*}[htb]\bc
\resizebox{\hsize}{!}{\includegraphics{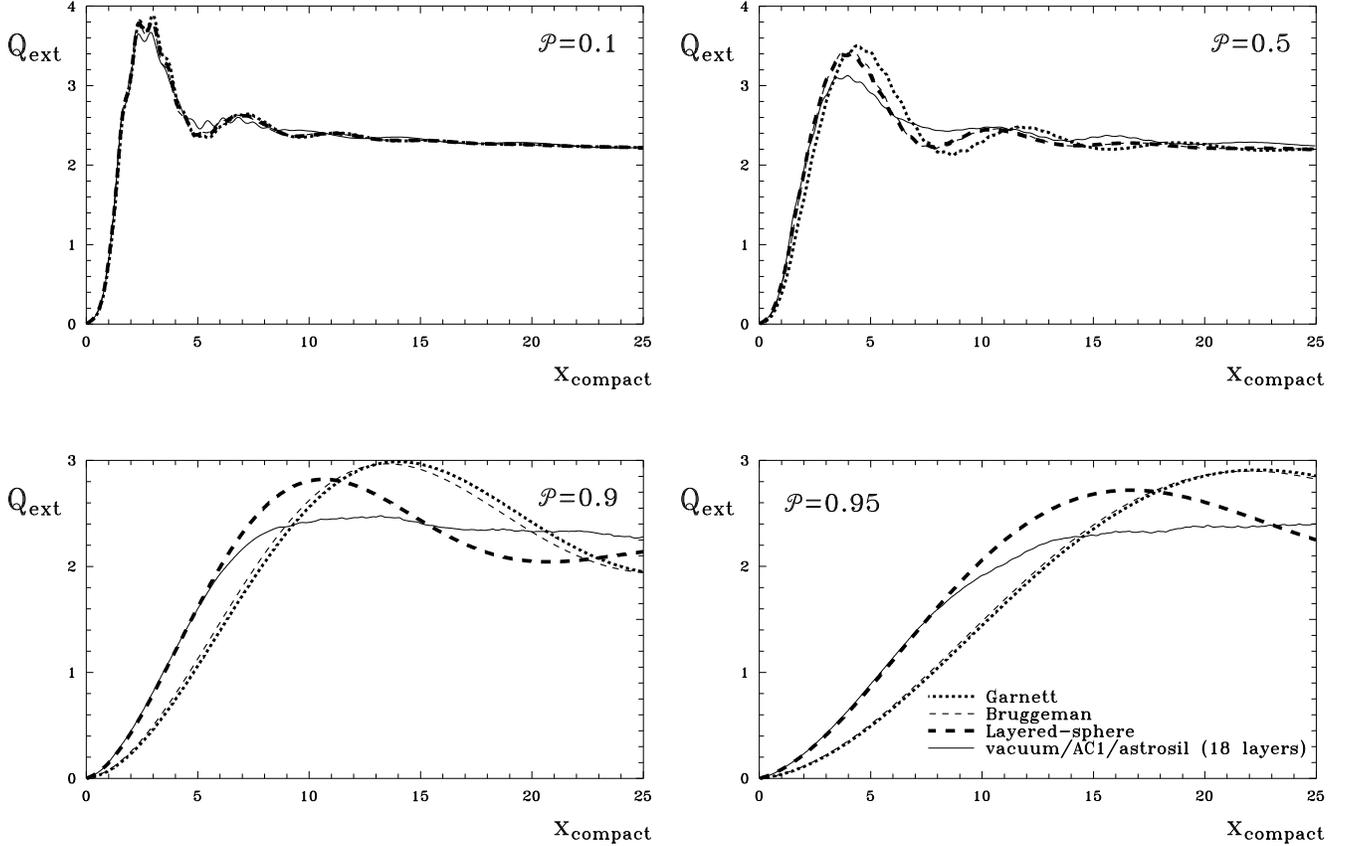}}
\caption{
   Size dependence of the efficiency factors
   calculated with the exact theory for multi-layered spheres
   and with the Mie theory using three different EMT rules.
   Multi-layered particles consist of 18 layers (6 shells).
   Each shell contains the same volume fractions of AC1 and
   astrosil and a varied fraction of vacuum.
   The cyclic order of the layers is indicated.
}\label{emt2}\ec\end{figure*}


Now let us discuss how different EMT rules can reproduce
the optical properties of multi-layered spheres.
Figure~\ref{emt1} (left panel)  shows the extinction efficiency factors computed
with the exact theory for 18 layered spheres (the order
of materials is vacuum/AC1/astrosil) and with the Mie theory  using
Garnett,\footnote{Vacuum was adopted as a matrix.}
Bruggeman, and layered-sphere mixing rules of
the EMT. The effective refractive indices are equal to
$m_{\rm eff}=1.496+0.060i$,
$m_{\rm eff}=1.541+0.081i$ and
$m_{\rm eff}=1.529+0.080i$
for Garnett, Bruggeman and  layered-sphere mixing rules, respectively.
Wiener's  maximum bound is $m_{\rm eff}=1.604+0.105i$.
Figure~\ref{emt1} (right panel) demonstrates the relative errors
for these EMTs.
It can be seen that the errors of the Bruggeman and layered-sphere rules
are  of several percent or better in the considered range of particle sizes.
The same is generally true for other
efficiency factors and albedo.
 An exception is the region after the first maximum of
the scattering efficiency factor and albedo ($x_{\rm porous} \approx 6 - 8$)
where the relative errors may reach up to 20\%. The largest errors
occur for the asymmetry factor, especially for small size parameters.
 The high accuracy of the layered-sphere rule in the case of
 very small particles sizes
is explained by the fact that it is based on the Rayleigh approximation.

 It should be noted that the order of materials playing a role,
for particles with several layers (see Sect.~\ref{eff}),
is also important for the  considered 18-layered particles.
For example, for spheres with another order of materials
(e.g., AC1/vacuum/astrosil) the interference maxima are stronger
than in the above-considered case (see Figs.~4, 5 and 8) and
as a consequence, the relative errors for the EMT rules
are larger, sometimes well exceeding 20\%.
 Other rules of the EMT behave like the Garnett and Bruggeman
rules, and we can conclude that despite the general condition of
the EMT applicability is not fulfilled for layered particles
-- ``inclusions'' (layers) are not small in comparison with
the wavelength of incident radiation -- most rules of the EMT
can reproduce the optical properties of layered
spheres of any size, if the number of layers is larger than 15~--~20.
This conclusion can be, however, affected by the porosity of particles.

Figure~\ref{emt2}  illustrates the applicability of different EMT rules
to multi-layered spheres of varying porosity.
The cases of other efficiency factors, albedo and asymmetry factor
are similar. The Figure demonstrates that the layered-sphere rule has
errors smaller than 10~--~20\% for any porosity and size of particles,
whereas other  rules provide unacceptable approximations
for large porosity (${\cal P} \ga 0.5$).
From Fig.~\ref{emt2} (lower panels)  one can see that the reason of
this advantage of the layered-sphere
rule is its accuracy in the Rayleigh domain ($x \rightarrow$ 0).
 Here again other rules of the EMT behave like the  Garnett
and Bruggeman ones, and it can be concluded that if the multi-layered
model provides a good approximation for particles with nearly
homogeneous distribution of several materials, the classical
EMT rules cannot be applied to strongly porous scatterers
of this kind.

\section{Wavelength dependence of extinction}\label{ext1}

\begin{figure*}[Htb!]\bc
\resizebox{\hsize}{!}{\includegraphics{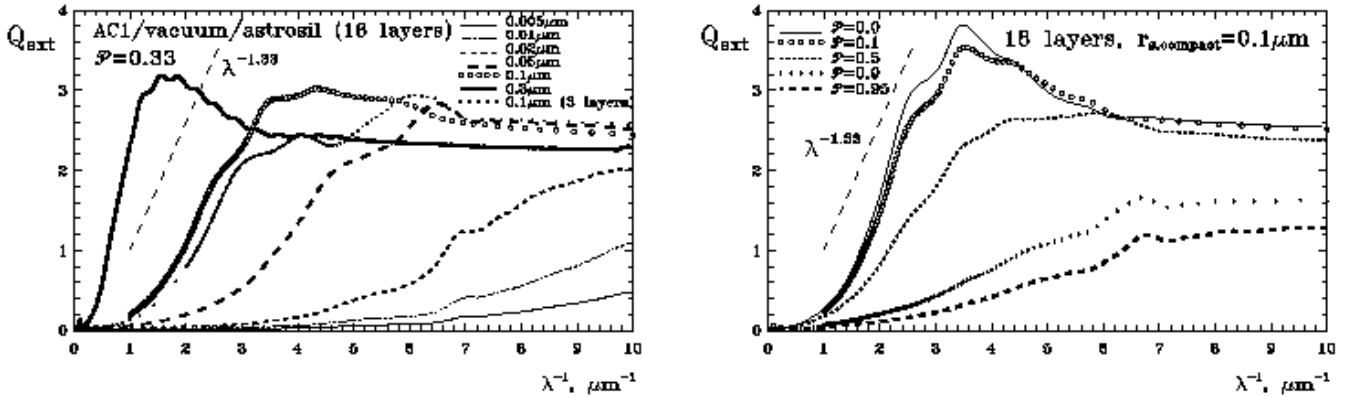}}
\caption{Wavelength dependence of the extinction efficiency factors
for multi-layered spherical particles. Left panel:
particles of different sizes,
each particle contains an equal fraction
of AC1,  astrosil and  vacuum separated.
The cyclic order of the layers is AC1/vacuum/astrosil.
Right panel: particles of the same mass but of different porosity.
}\label{ext_w}\ec\end{figure*}

As it is well known,
the wavelength dependence of interstellar extinction $A(\lambda)$
is completely determined by the wavelength dependence of the
extinction efficiencies $Q_{\rm ext}(\lambda)$.
This quantity is shown in Fig.~\ref{ext_w} for multi-layered particles
of different sizes and porosity.
The average interstellar extinction curve in the visible--near UV
($1 \,\mkm^{-1} \leq\lambda^{-1} \leq 3 \,\mkm^{-1}$)
can be approximated by the power law $A(\lambda) \propto \lambda^{-1.33}$
(see discussion in Voshchinnikov~\cite{v02}).
This dependence is plotted in Fig.~\ref{ext_w} as a dashed segment.
 From the Figure one can conclude that  extinction
depends on both the particle size and its chemical composition, i.e.
the volume averaged refractive index.
This mean refractive index is maximal for compact grains.
Note that  particles with 3 layers and  those with 18 layers
produce similar dependences $Q_{\rm ext}(\lambda)$
in the visual part of the spectrum (Fig.~\ref{ext_w}, left panel).

As it has been mentioned many times (e.g., Greenberg \cite{g78}),
a comparable extinction occurs if the product of the typical
particle size $\langle r \rangle$ with the refractive index is constant
\be
\langle r \rangle  \, |m-1| \approx {\rm const.}
\label{mr}\ee
However, this conclusion breaks down if one considers
very porous particles.
The average refractive index of
particles with larger fraction of vacuum shown in Fig.~\ref{ext_w}
(right panel) is closer to 1 but their radii are larger
(e.g., from Eq.~(\ref{xpor}) follows that
$r_{\rm s}=0.27 \,\mkm$ if ${\cal P}=0.95$ and $r_{\rm s,\,cmpact}=0.1 \,\mkm$).
Despite of the equal amount of solid material in  particles, the wavelength
dependence of extinction becomes flatter. The analogous behaviour
is typical of particles of other masses (i.e., for
compact spheres of other radii), and, therefore, the reconstruction
of the observed interstellar extinction curve
with very porous grains only should will be rather difficult.

\section{Radiation pressure}\label{pr}

It is generally accepted that the mass loss in evolved stars
is controlled by the radiation pressure on dust grains.
The radiation pressure also affects the motion of  interplanetary and
interstellar grains. The capacity of a star
with the effective temperature $T_\star$,
radius $R_\star$ and mass $M_\star$
to move a grain with the radius $r_{\rm s}$
is characterized by the sweeping efficiency
(the ratio of the radiation pressure force to the gravitational one,
see, e.g., Voshchinnikov \& Il'in \cite{vi83})
\be
\beta \equiv \frac{F_{\rm pr}}{F_{\rm g}} =
    \frac{\sigma}{c \, G}\, \frac{R^2_\star T^4_\star}{M_\star}\,
    \frac{\pi r^2_{\rm s} \overline{Q}_{\rm pr}}{\rho_{\rm d} V}.
    \label{be}
\ee
Here, $\sigma$ is the Stephan-Boltzmann constant,
$c$ the speed of light, $G$ the gravitational constant,
$V$ the particle volume, $\rho_{\rm d}$ the material density,
$\overline{Q}_{\rm pr}$ the Planck mean radiation pressure efficiency.

\begin{figure}[htb]\bc
\resizebox{\hsize}{!}{\includegraphics{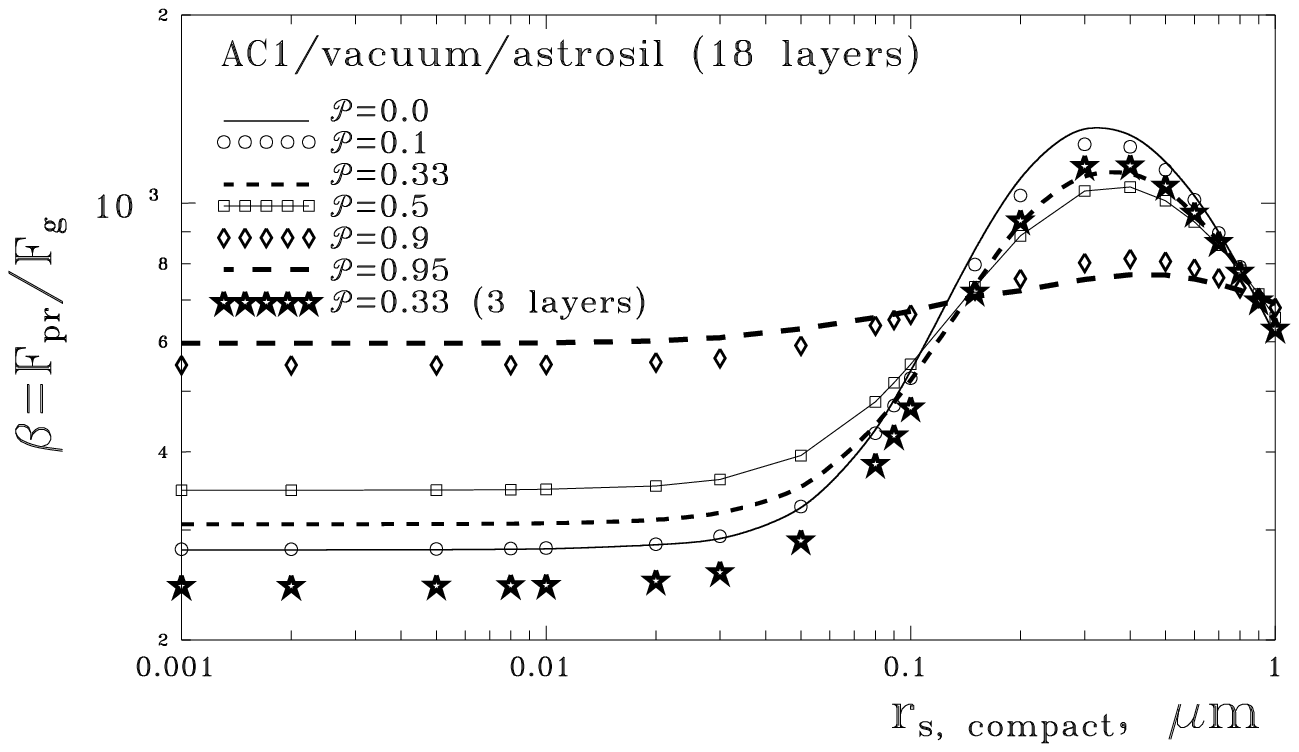}}
\caption{Size dependence  of the sweeping efficiency
for multi-layered spheres.
The cyclic order of the layers is AC1/vacuum/astrosil.
The particles are located near a star with the effective temperature
$T_\star=2500$\,K, radius $R_\star=300\,R_\odot$ and mass
$M_\star=2\,M_\odot$.
}\label{betpr}\ec\end{figure}

Figure~\ref{betpr} shows the size dependence  of $\beta$
for multi-layered porous and compact spheres.
The radius of porous particles is calculated from  Eq.~(\ref{xpor})
and their density is given in Table~\ref{t1}.
As follows from Fig.~\ref{betpr}, an increase of particle porosity
results in the increase of the sweeping efficiency for small
particles ($r_{\rm s,\,compact} \la 0.1\,\mkm$) and the decrease
of $\beta$ for large particles ($r_{\rm s,\,compact} \ga 0.1\,\mkm$).
For very porous grains, the dependence $\beta(r_{\rm s})$ is flat, i.e.
almost size independent. So, the motion of such particles
will be determined by the  drag force due to collisions  with
 gas particles.

Previous calculations of radiation pressure on porous particles
have been made,  using  Mie theory and the Bruggeman mixing rule
(Il'in \& Krivova \cite{ik00}) and the DDA
for fluffy dust aggregates
(Kozasa et al. \cite{kbm92}, Kimura et al. \cite{kom02}).
The behaviour of the sweeping efficiency for aggregates
of increasing porosity (the DDA calculations)
 seems to be similar to that found by us
while the Mie--Bruggeman theory  leads to the results
contradicting those presented in Fig.~\ref{betpr}.

\section{Scattered radiation}\label{sca}
\subsection{Albedo and asymmetry parameter}

\begin{figure*}\bc
\resizebox{\hsize}{!}{\includegraphics{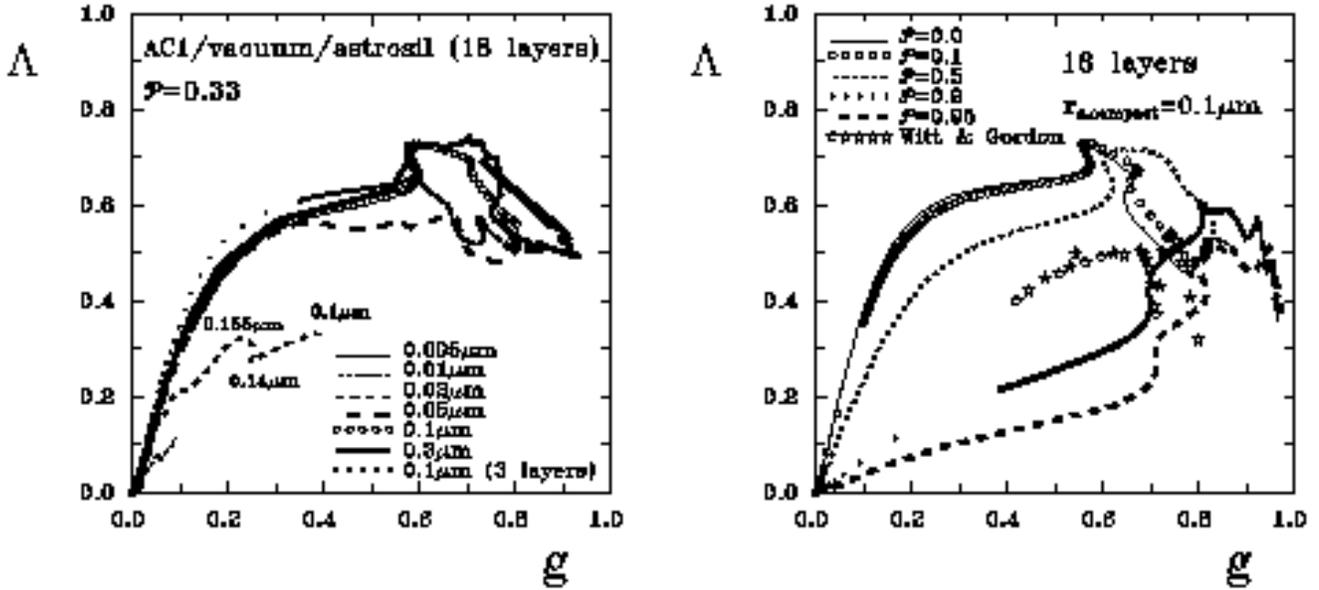}}
\caption{
Albedo dependence on the asymmetry parameter for multi-layered spherical particles.
The particle parameters are the same as in Fig.~\ref{ext_w}.
All curves start at the wavelength $\lambda=20\,\mkm$,
where $\Lambda \simeq g \simeq 0$, and finish at $\lambda=0.1\,\mkm$.
The turning points for the particle with $r_{\rm s, \, compact} = 0.02\,\mkm$
are marked at  the left panel.
At the right panel, the stars present the summary results of determination of
$\Lambda$ and $g$   in the wavelength
range $\lambda=0.1-3.0\,\mkm$ for our Galaxy (Witt \& Gordon,~\protect\cite{wg00},
Table~1; see also discussion in the text).
}\label{lg}\ec
\end{figure*}
\begin{figure*}\bc
\resizebox{\hsize}{!}{\includegraphics{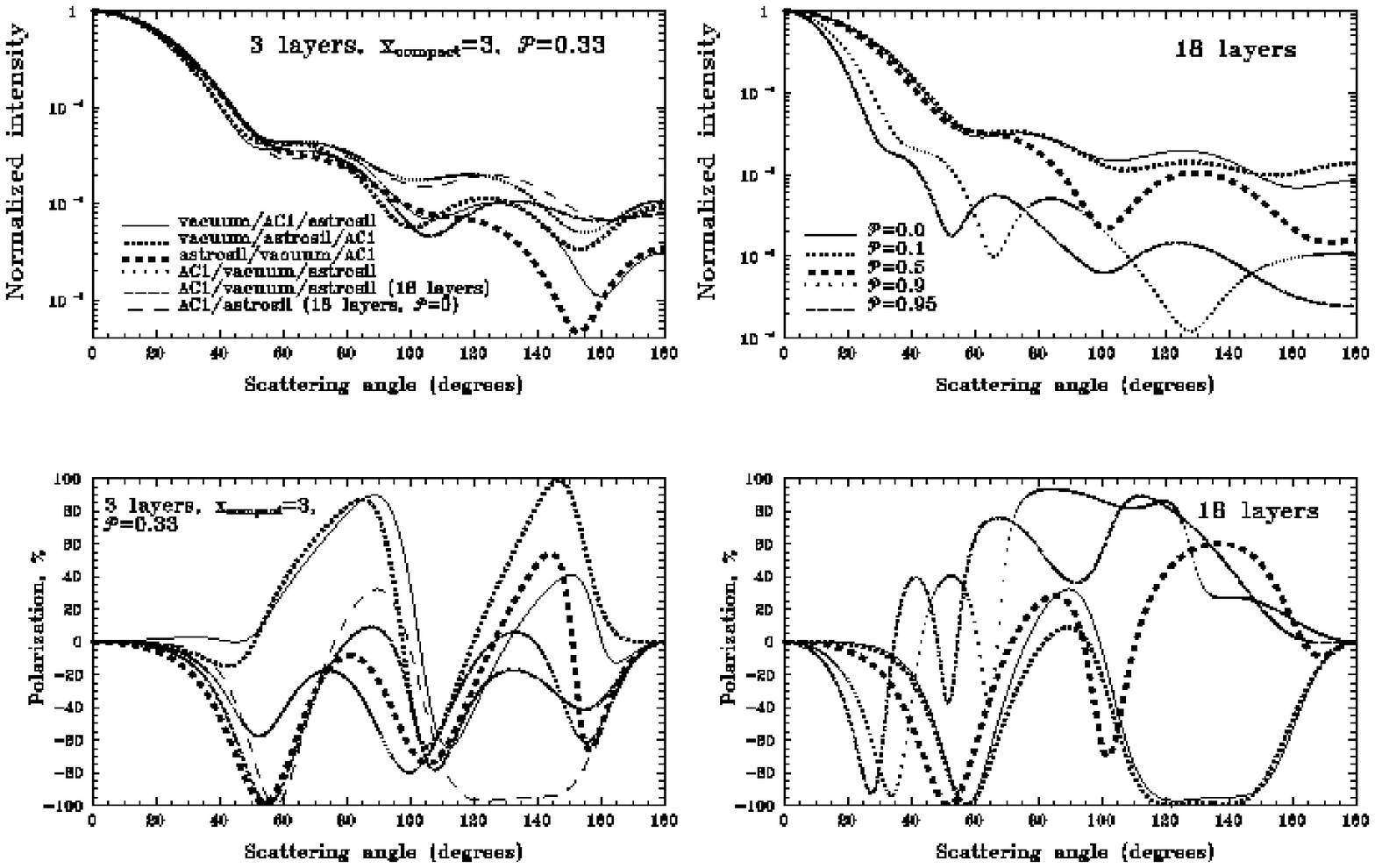}}
\caption{
Normalized intensity  and polarization of scattered radiation
for multi-layered spheres. Left panels:
particles of the same size ($x_{\rm compact}=3$)
but different cyclic order of the layers.
Right panels: particles of the same mass but  different porosity.
the cyclic order of the layers is AC1/vacuum/astrosil.
}\label{iipp}\ec
\end{figure*}

In modelling of light scattering in dusty objects,
the Henyey \& Greenstein~(\cite{hg41}) phase function is often used.
It is parameterized by the asymmetry parameter $g$.
Then albedo $\Lambda$ and $g$ are utilized as parameters for
radiative transfer problems irrelevant to  the actual properties of dust
grains.

The standard behaviour of albedo and asymmetry parameter
is as follows: $\Lambda \ll 1$ and $g \simeq 0$ for small size
parameters (small sizes or large wavelengths), then
$\Lambda$ and $g$ grow with increasing
particle size or decreasing   wavelength
and reach the asymptotic values for very large size parameters.
The behaviour of $\Lambda$ vs.
$g$ is plotted in Fig.~\ref{lg} for porous and compact particles.
It is important to keep in mind that the albedo and asymmetry parameter
cannot be determined from observations separately,
but only together (as a combination). Therefore,
models with one fixed parameter and the other varying
often make little physical sense.
We only mention  the paper of Witt \& Gordon~(\cite{wg00})
who compiled the results of the determination of $\Lambda$ and $g$
for galactic dust from 13 publications. These data are  mainly based
on the previous modelling of scattered light from reflection nebulae
and diffuse galactic light in the visual and UV parts of spectrum.
Witt \& Gordon visually averaged the wavelength dependencies
of albedo and asymmetry parameter and shifted the data for $\Lambda$
to reach agreement with the dust model of Kim et al.~(\cite{kmh94}).
As a result, the dependencies  $\Lambda(\lambda)$ and $g(\lambda)$
for wavelengths $\lambda=0.1-3.0\,\mkm$
were published in tabular and graphical forms.
These data are shown in  Fig.~\ref{lg} (right panel).
It can be seen that depending on the particle size and wavelength of
radiation the position of any particle in the  $\Lambda - g$ plane is well fixed.
{\it Some pairs of parameters {\rm (}$\Lambda$, $g${\rm )}
have no correspondence to any particle.}
In particular, in order to fill in the right lower corner
in such diagrams, the particles should be very porous (${\cal P} \ga 0.7$).
Note that neither Witt \& Gordon~(\cite{wg00}),
nor Kim et al.~(\cite{kmh94}) considered  porous grains.
The theoretical constraints on
the albedo and asymmetry parameter were also discussed by Chlewicki
\&  Greenberg~(\cite{cg84}) who showed that the results of
some models could not be realized by the optics of small particles.

\subsection{Intensity and polarization}

The angular dependence of the  intensity of scattered radiation is mainly
determined by the particle size parameter (van de Hulst~\cite{vdh},
Bohren \& Huffman~\cite{bh83}).
This is well seen when the normalized intensity is considered.
Variations in the order of layers have a small effect
on the intensity in the case of scattering in the forward
hemisphere (the scattering angles $\Theta <90\degr$, Fig.~\ref{iipp},
left upper panel). In the backward hemisphere ($\Theta > 90\degr$),
the dependencies of $I(\Theta)$ for particles consisting of 18 layers
being compact or having intermediate porosity
(${\cal P}=0.33$) are similar. Note that
the order of layers becomes important
for three-layered particles at $\Theta \ga 120\degr$.

The influence of the particle porosity on the
normalized intensity is shown in
Fig.~\ref{iipp} (right upper panel). The growth of porosity is accompanied by
an increase of the particle size because the particle mass is fixed.
In the case of very large values of ${\cal P}$, the intensity strongly
grows  for nearly  forward directions ($\Theta \la 40\degr$), which looks
similar to the behaviour of intensity for large transparent spheres.

In contrast to the intensity, the polarization of the scattered radiation
strongly depends on  the order of layers in the particle
as well as its  porosity (see Fig.~\ref{iipp}, lower panels).
Very large differences in polarization occur at almost all scattering angles.
The change of the order of layers and
the porosity can result in a change of the sign of polarization.
In the case of the three-layered particles (Fig.~\ref{iipp}, left lower panel),
the polarization is mainly positive if vacuum is in the core and
the polarization is mainly negative if vacuum forms an intermediate
layer. At large scattering angles ($\Theta \ga 110\degr$),
the negative polarization  transforms to the positive one when
${\cal P}$ increases (Fig.~\ref{iipp}, right lower panel).

\section{Infrared radiation}\label{irr}
\subsection{Temperature}\label{temp}

\begin{figure}\bc
\resizebox{\hsize}{!}{\includegraphics{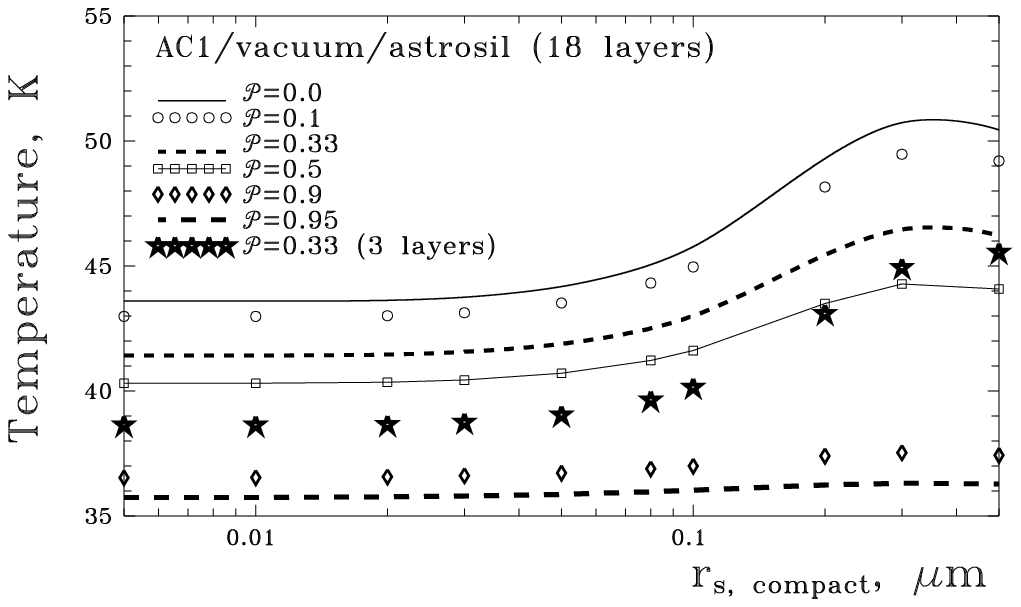}}
\caption{Size dependence  of the temperature
for multi-layered spheres.
The cyclic order of the layers is AC1/vacuum/astrosil.
The particles are located at a distance of
10$^4\,R_\star$ from a star with the effective temperature
$T_\star=2500$\,K.
The radius of porous particles is calculated from Eq.~(\ref{xpor}).
}\label{td}\ec\end{figure}

The commonly considered equilibrium temperature
of cosmic grains is
the result of  absorption of the UV and visual stellar
photons and re-emission of IR photons.
Figure~\ref{td} shows the temperature of multi-layered particles,
depending on their size.
The results were calculated for particles located at a distance of
10$^4\,R_\star$ from a star with the effective temperature
$T_\star=2500$\,K. In all cases, the increase of the vacuum volume fraction
causes a decrease of the grain temperature if the  amount of the
solid material is kept constant.
This result holds for particles of larger sizes
located closer to (or farther from) the star and for other values
of $T_\star$. If the  materials are well mixed (the number
of shells $\ga 3-4$), the temperature  drops when
the porosity grows. Such a behaviour contradicts the results
of Greenberg \& Hage~(\cite{grha91}) who found an increase of
temperature with grain porosity (see their Fig.~4).
The discrepancy is explained by the impossibility of the
the method used by Greenberg \& Hage to treat very porous particles
(see also Sect.~\ref{ir_b}).

If the  materials are badly mixed, the grain temperature
is determined by the position of the most absorbing component
(amorphous carbon) in the particle. If AC1 is in the core
(``stars'' in Fig.~\ref{td}), the  temperature
will be 3~--~6\,K lower  than in  the case when AC1
forms an intermediate or outermost layer.

\subsection{Infrared features}\label{ir_b}

\begin{figure}\bc
\resizebox{\hsize}{!}{\includegraphics{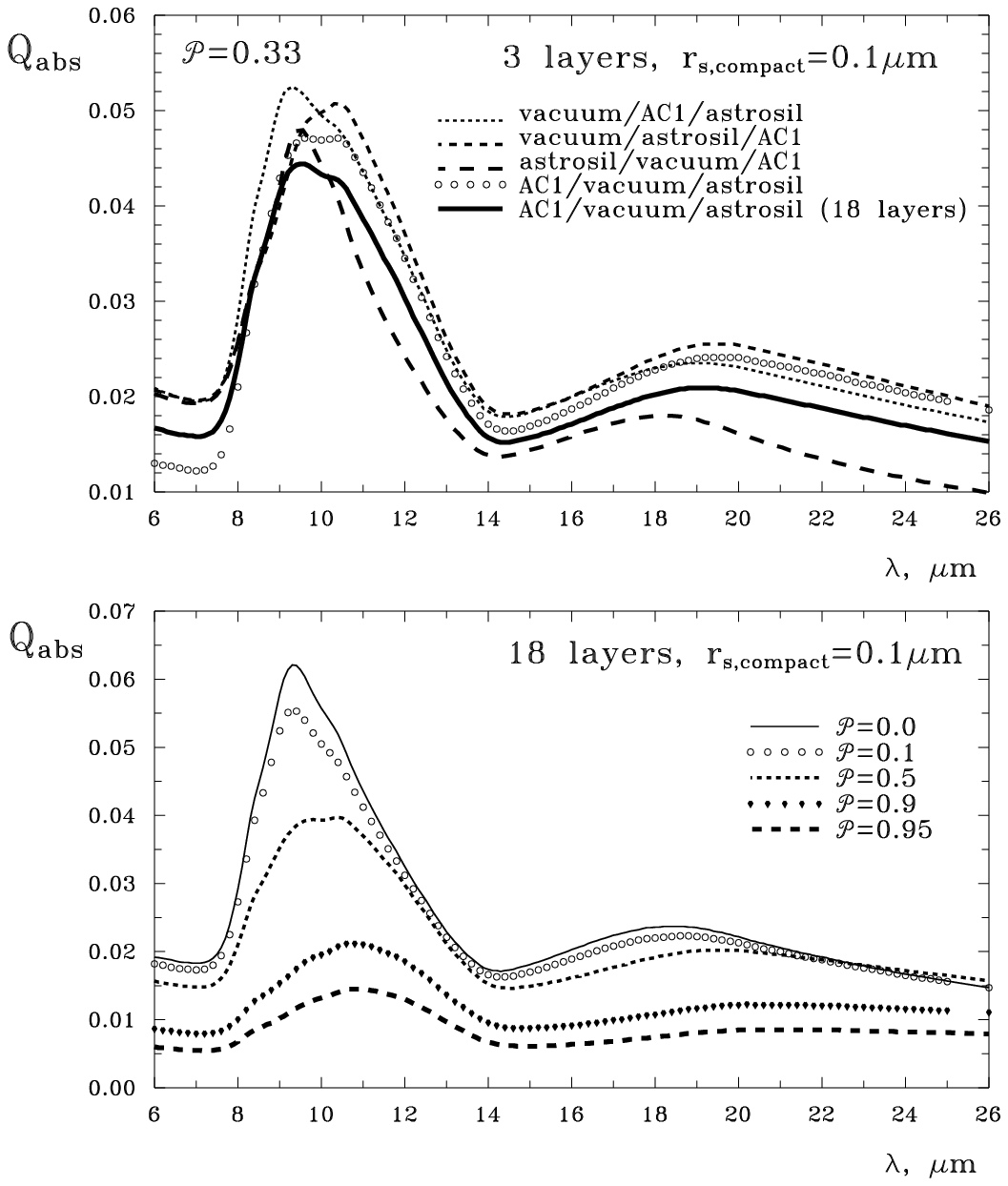}}
\caption{Wavelength dependence  of the absorption efficiency factors
for multi-layered spherical particles.
}\label{abs01}\ec\end{figure}

As it is well known, the shape of IR dusty features is a good indicator
of particle size and chemical composition.  With  increasing size,
a  feature becomes wider and wider and fades away.
For example, in the case of compact spherical grains of astrosil,
the 10~$\mkm$ and  18~$\mkm$  features disappear when the grain radius
exceeds $\sim 2-3 \,\mkm$.
The small scale structure of features
are usually attributed to the variations of components of grain
material (e.g., changes of the ratio of magnesium
to iron in silicates) or the modification of
the material  state (amorphous/crystalline).

In Fig.~\ref{abs01} we compare the wavelength dependence of
the absorption efficiency factors for particles of the same mass but different
structure.
The upper panel shows the results for particles of the same porosity
consisting of one shell (three layers). In this case,
the  central position and the width of dust features
depend on the order of layers. Larger changes occur for the red wing.
Note that the normalized factors are very similar for particles
of different sizes (if $r_{\rm s, \, compact} \la 0.5\,\mkm$)
and vary only with particle structure.
As in the previous cases, with the increase of the number of layers
the results converge to some limit (thick solid curve in Fig.~\ref{abs01}).

In the lower part of Fig.~\ref{abs01}, variations of $Q_{\rm abs}$
with the particle porosity are plotted.
The growth of ${\cal P}$ causes of the shift the center
of the feature to longer wavelengths and its broadening.
For  particles with ${\cal P}=0.95$,
the 10~$\mkm$  feature transforms into a plateau while
the 18~$\mkm$  feature disappears.
This tendency is opposite to the results
shown in Fig.~7 of Hage \& Greenberg~(\cite{hagr90}) who found
that the higher the porosity, the sharper the silicate emission becomes,
but agrees with the DDA calculations of Henning \& Stognienko~(\cite{hs93}).

\begin{figure}\bc
\resizebox{\hsize}{!}{\includegraphics{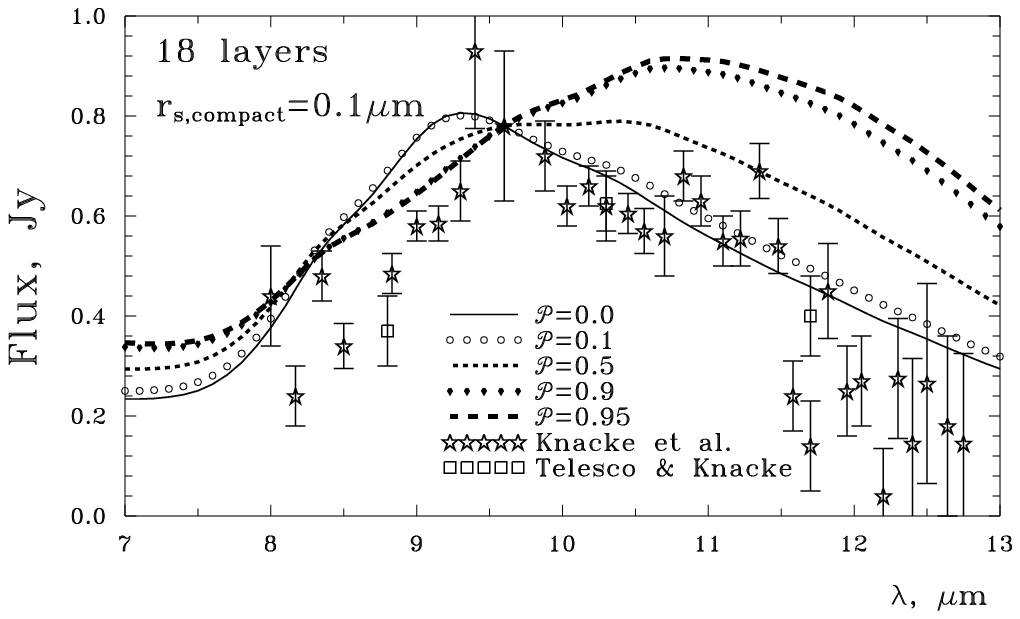}}
\caption[]
{Emission in the disc around the star $\bet$  in the region of silicate
10\,$\mkm$ band.
Stars and squares are the observations of Knacke et al.~(\cite{kn93})
and of Telesco \& Knacke~(\cite{tk91}). The curves present the results
of calculations for multi-layered particles
of radius $r_{\rm s, \, compact}=0.1 \,\mkm$
as shown in Fig.~\ref{abs01} but normalized at $\lambda=9.6\,\mkm$.
}\label{bet}\ec\end{figure}

Their  model was used later by Li \& Greenberg~(\cite{li:gre98}) for
the explanation of the 10~$\mkm$ emission feature in the
disc of $\bet$. The best fitting was obtained for highly
porous particles with ${\cal P} \approx 0.95$.
In order to estimate the validity of this conclusion,
we plotted in Fig.~\ref{bet} our data from Fig.~\ref{abs01} in a normalized
manner together with observations of $\bet$ made by
Knacke et al.~(\cite{kn93}) and Telesco \& Knacke~(\cite{tk91}).
As follows from Fig.~\ref{bet}, the observed shape of the silicate feature
is better reproduced if particles are compact or the porosity is small.

\subsection{Dust opacities}\label{opa}

The dust opacity or
the mass absorption coefficient of a grain material $\kappa(\lambda)$
enters directly in the expression for
the dust mass of an object $M_{\rm d}$ which is determined from
IR observations
\be
M_{\rm d} =  \frac{F_{\rm IR}(\lambda) D^2}{\kappa(\lambda) B_\lambda(T_{\rm d})}.
        \label{m}
\ee
Here, $F_{\rm IR}(\lambda)$ is  the observed flux,
$D$ the distance to the object, $B_\lambda(T_{\rm d})$ the Planck function,
$T_{\rm d}$ the dust temperature.
The quantity $\kappa(\lambda)$ depends on
the particle volume $V$,  the material density $\rho_{\rm d}$
and the extinction cross-section $C_{\rm ext}$
\bea
\kappa(\lambda) = \frac{C_{\rm ext}}{\rho_{\rm d} V}  \approx
     \frac{3}{\rho_{\rm d}} \, \left(\frac{2 \pi}{\lambda}\right) \, {\rm Im} \left\{\frac{\alpha}{V} \right\} =
\nonumber \\
\,\,\,\,\,\,\,    \frac{3}{\rho_{\rm d}} \, \left(\frac{2 \pi}{\lambda}\right) \,
 {\rm Im} \left\{\frac{\ve_{\rm eff}-1}{\ve_{\rm eff}+2} \right\}.
        \label{kap}
        \eea
At long wavelengths the scattering can be neglected
and $C_{\rm ext} \approx C_{\rm abs}$.
We also can evaluate $C_{\rm abs}$ in the Rayleigh approximation,
then the mass absorption coefficient will not depend on the particle size
as follows from the right part of Eq.~(\ref{kap}).
The effective dielectric permittivity in Eq.~(\ref{kap}) can be found from
the layered-sphere rule of the EMT (see Eqs.~(\ref{ls1}), (\ref{ls2})).

\begin{table}[htb]
\caption[]{Opacities at $\lambda = 1$\,mm of
multi-layered and compact spheres of the same mass.}    \label{t1}
\bc\begin{tabular}{lccc}
\hline
\noalign{\smallskip}
Particle & ${\cal P}$ &  $\rho_{\rm d}$, g/cm$^3$  & $\kappa$, cm$^2$/g \\
\noalign{\smallskip}
\hline
\noalign{\smallskip}
18 layers: AC1$^\ast$/astrosil$^{\ast\ast}$  & 0.00 & 2.58 &   ~0.948   \\
\noalign{\smallskip}
3 layers: vac/AC1/astrosil            & 0.33 & 1.72   &   2.84  \\
\phantom{3 layers: }vac/astrosil/AC1  & 0.33 & 1.72   &   3.63   \\
\phantom{3 layers: }astrosil/vac/AC1  & 0.33 & 1.72   &   3.41   \\
\phantom{3 layers: }AC1/vac/astrosil  & 0.33 & 1.72   &   1.37   \\
\noalign{\smallskip}
18 layers: vac/AC1/astrosil           & 0.33 & 1.72   &   2.25 \\
\phantom{18 layers: }vac/astrosil/AC1 & 0.33 & 1.72   &   2.22  \\
\phantom{18 layers: }astrosil/vac/AC1 & 0.33 & 1.72   &   2.39  \\
\phantom{18 layers: }AC1/vac/astrosil & 0.33 & 1.72   &   1.96  \\
\noalign{\smallskip}
18 layers: AC1/vac/astrosil           & 0.10 & 2.32   &   1.55  \\
\phantom{18 layers:AC1/vac/astrosil}  & 0.30 & 1.73   &   2.20  \\
\phantom{18 layers:AC1/vac/astrosil}  & 0.50 & 1.29   &   2.57  \\
\phantom{18 layers:AC1/vac/astrosil}  & 0.70 & ~0.772 & 4.04 \\
\phantom{18 layers:AC1/vac/astrosil}  & 0.90 & ~0.258 & 8.12 \\
\phantom{18 layers:AC1/vac/astrosil}  & 0.95 & ~0.129 & $\!\!\!$ 10.4  \\
\noalign{\smallskip}
\hline
\noalign{\smallskip}
\end{tabular}\ec
\noi $^{\ast}$ $m(1 {\rm mm})=2.93+0.276i$, $\rho_{\rm d}=1.85$\,g/cm$^3$ \\ 
\noi $^{\ast\ast}$ $m(1 {\rm mm})=3.43+0.050i$, $\rho_{\rm d}=3.3$\,g/cm$^3$ \\
\end{table}

Extensive study of the dependencies of the mass absorption coefficients
on the material properties and grain shape is
summarized in the paper of Henning~(\cite{h96}) where,
in particular, it is noted that the opacities at 1~mm are considerably
larger for non-spherical particles in comparison with spheres
(see also Ossenkopf \& Henning \cite{oh94}).
We find that the  similar effect yields  the inclusion of vacuum as a component
of composite particles. This follows  from Table~\ref{t1} where
the opacities at $\lambda = 1$\,mm are given for spheres of the same mass.
It can be seen that the values of $\kappa$ are generally larger for particles
with larger fraction of vacuum and in the case of amorphous carbon
as an outer layer. The last effect is reduced when the number of layers increases.
Note that the growth of the opacity with particle porosity is related
to the decrease of the particle density $\rho_{\rm d}$ which is calculated as the
volume-average quantity.
Because the mass of dust in an object is proportional to $\rho_{\rm d}$
(see Eqs.~(\ref{m}) and (\ref{kap})), the assumption on porous grains
leads to smaller masses.

\section{Multi-layered grains and cosmic abundances}\label{abun}
\subsection{Interstellar   abundances and extinction}\label{ab_ext}

The basic requirement for any model of interstellar dust is the
explanation of the observed extinction law taking into account
the dust-phase abundances of elements in the interstellar medium.
These abundances are obtained as the difference between the
observed gas-phase abundances and the cosmic reference ones.
However, the cosmic abundances are not yet
conclusively established and usually this causes a problem.
For many years, the solar abundances were used as the reference ones,
until the photospheres of the early-type stars were found
not to be so rich in heavy elements as the Sun was (Snow \& Witt \cite{sw96}).
Such a revision of the reference abundances caused the
so-called ``carbon crisis'': abundances of the most important dust-forming
elements (C, O, Mg, Si, Fe) required by the current dust models are
greater than the  abundances available in the solid phase of
the interstellar medium.
In the modelling of the interstellar extinction the solar abundances are
still utilized (e.g., Weingartner \& Draine \cite{weid01},
Clayton et al. \cite{cletal03}), although   high-resolution
\mbox{X-ray} spectroscopy has already given  direct evidence of the subsolar
interstellar abundance of oxygen.
Using the Chandra high resolution spectrum of the object Cyg~X-2,
Takei et al.~(\cite{takei02}) estimated the
dust-phase (179 ppm\footnote{particles per million hydrogen atoms})
and total (579 ppm) abundances of oxygen.
The latter is $\sim 68\%$ of the solar one (851 ppm).

Our model of multi-layered porous particles is applied
for the explanation of the absolute extinction in the direction to two stars.
The model could provide more extinction per unit of mass than models with
compact particles  (see Fig.~\ref{fvac}, lower panel).
There were many unsuccessful attempts to resolve the crisis.
Here we analyse the principal possibility to do that and
to enlarge the extinction per unit of mass in order to
minimize the amount of solid phase substance.
We considered several materials as components of composite grains.
Among the carbonaceous species,
the amorphous carbon in the form of Be1 (Rouleau \& Martin \cite{rm91})
was found to produce the largest
extinction. Also the extinction of iron oxides strongly increases
with the growth of porosity.

\begin{figure}[htb]
\begin{center}
\resizebox{\hsize}{!}{\includegraphics{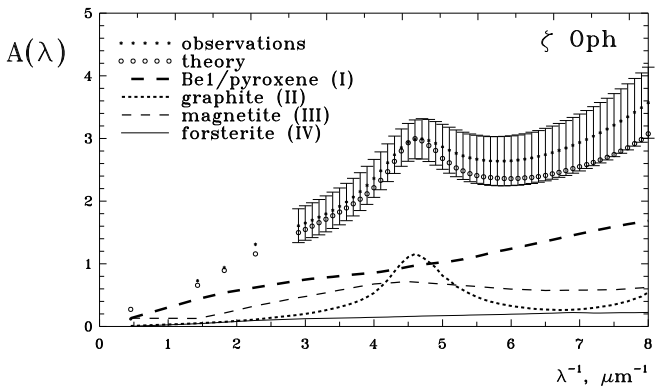}}
\caption
{Observed and calculated extinction in the direction to the star
$\zeta$ Oph.
The errors of the observations are the result of a parameterization
of the observations (see Fitzpatrick \&  Massa \cite{fm90}).
The contribution to the theoretical extinction from different components
is shown.}
\label{zeta}\end{center}\end{figure}

\begin{figure}[htb]
\begin{center}
\resizebox{\hsize}{!}{\includegraphics{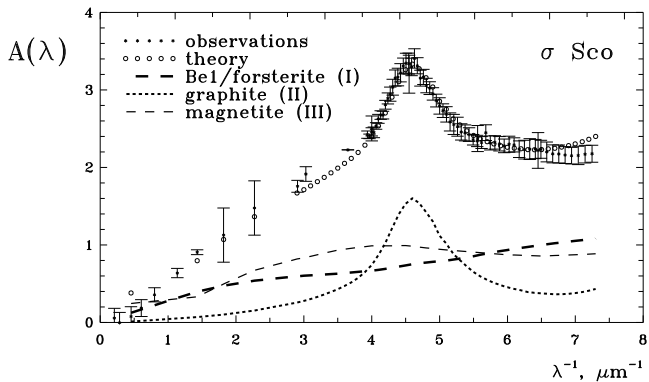}}
\caption
{The same as in Fig.~\ref{zeta} but now for the star $\sigma$ Sco.
The observational data were taken from  Wegner~(\cite{ww02}).
}\label{sigma}\end{center}\end{figure}

We fitted the observed extinction  toward the stars
$\zeta$~Oph (Fig.~\ref{zeta}) and $\sigma$ Sco (Fig.~\ref{sigma}).
The models consist of three or four grain populations.
\begin{itemize}
\item (I). Porous composite (multi-layered) particles
(Be1 --- 5\%,
pyroxene, Fe$_{0.5}$Mg$_{0.5}$SiO$_3$ --- 5\% for $\zeta$ Oph or
forsterite, Mg$_2$SiO$_4$ --- 5\% for $\sigma$ Sco and
vacuum --- 90\%)
with the power-law size distribution having an exponential decay.
The lower/upper cut-off in the size distribution is
$0.005\,\mkm$/$0.25\,\mkm$ and $0.05\,\mkm$/$0.35\,\mkm$ in the case
of $\zeta$~Oph and $\sigma$ Sco, respectively.
\item (II). Small compact graphite grains
with the narrow power-law size distribution ($r_{\rm s}\approx 0.01 - 0.02\,\mkm$).
\item (III). Porous composite grains of magnetite (Fe$_3$O$_4$ --- 2\%, vacuum --- 98\%)
with the power-law size distribution.
The lower/upper cut-off in the size distribution is
$0.005\,\mkm$/$0.25\,\mkm$ and $0.09\,\mkm$/$0.50\,\mkm$ in the case
of $\zeta$~Oph and $\sigma$ Sco, respectively.
\item (IV). Compact grains of forsterite (Mg$_2$SiO$_4$)
with the power-law size distribution (only for $\zeta$ Oph,
$r_{\rm s,\,min}=0.005\,\mkm$, $r_{\rm s,\,max}=0.25\,\mkm$).
\end{itemize}
The contributions from different components to the calculated extinction
are shown in Figs.~\ref{zeta} and \ref{sigma}.

\begin{table}[htb]
\bc
\caption[]{Interstellar ``cosmic'' and dust-phase  abundances (in ppm)}\label{ida}
\begin{tabular}{cccccc}
\hline
\noalign{\smallskip}
Element & ``cosmic" & \multicolumn{2}{c} {$\zeta$ Oph} &\multicolumn{2}{c} {$\sigma$ Sco}\\
 & abundance& obs & model & obs & model \\
\noalign{\smallskip}
\hline
\noalign{\smallskip}
~C& 214   & ~79  &  195  &     145        & 121    \\
~O& 457   & 126  &  128  &     ~85        & ~66    \\
Mg$^\ast$& ~25   & ~23  &  ~25  &     ~22        & ~~~15.5 \\
Si& ~~18.6& ~17  &  ~30  &  ~~16.8        & ~~~7.5  \\
Fe& ~27   & ~27  &  ~34  & ~~~26.7        & ~~~26.4  \\
\noalign{\smallskip}\hline
\end{tabular}\ec
\noi $^\ast$ The abundance of Mg was recalculated with the new oscillator
strengths from Fitzpatrick~(\cite{fp97}).

\end{table}

Table~\ref{ida} gives the reference ``cosmic'' abundances
of five dust-forming elements according to Snow \& Witt~(\cite{sw96})
as well as the observed and model dust abundances.

The dust-phase abundances in the line of sight to the star
$\zeta$~Oph (HD~149757) were taken from
Table~2 of Snow \& Witt~(\cite{sw96}).
In our calculations, we adopted for $\zeta$~Oph: the total extinction
$A_{\rm V}=0\fm94^($\footnote{This value was obtained from the
relation $A_{\rm V}= 1.12 \, E({\rm V-K})$ (Voshchinnikov  \& Il'in \cite{vi87})
and colour excess  $E({\rm V-K})=0\fm84$ (Serkowski et al. \cite{smf75}).}$^)$,
colour excess $E({\rm B}-{\rm V})=0\fm32$ and the
total hydrogen column density $N({\rm H})=1.35\,10^{21}\,{\rm cm}^{-2}$
(Savage \&  Sembach, \cite{ss96}).
The extinction curve was reproduced according to the parameterization
of Fitzpatrick \&  Massa~(\cite{fm90}).

For $\sigma$ Sco (HD~147165), we used the extinction curve,
colour excess $E({\rm B}-{\rm V})=0\fm35$ and
the total extinction $A_{\rm V}=1\fm13$ according to Wegner~(\cite{ww02}).
The hydrogen column density
$N({\rm H})=2.46 \, 10^{21}\, {\rm cm}^{-2}$ was adopted from
Zubko et al.~(\cite{zkw96}). The gas-phase abundances were taken from
Allen et al.~(\cite{asj90}).

The dust-phase abundances required by the model  are larger
than the observed ones in the direction to $\zeta$~Oph and
smaller than the observed abundances in the direction to $\sigma$ Sco.
Note that for this star the required amount of C and Si in dust
grains is the lowest in  comparison with  previous
modelling. This is a result of the use of highly porous particles
which gives the considerable extinction and simultaneously
allows one to ``save'' the material.
For example, the extinction model of $\sigma$~Sco with compact grains
presented by Zubko et al.~(\cite{zkw96})
requires 240~--~260 ppm of C and 20~--~30 ppm of Si
and the model of Clayton et al.~(\cite{cletal03})
needs 155 ppm of C and 36 ppm of Si
(cf. 121 ppm and 7.5 ppm from Table~\ref{ida}).

Evidently, it remains an open question how to explain the observed
values of extinction of different stars with a reduced amount of heavy
elements (i.e. how to increase the extinction to volume ratio).
It seems that some enhancement of this ratio can be expected if one takes
into account a possible non-sphericity of
dust grains, but it is difficult to beleive that this
improvement will completely resolve the problem
(see Fig.~29 in Voshchinnikov~\cite{v02}).
It is also hardly probable that larger extinction values are
provided by an unknown material. The possible way to overcome
the carbon crisis may be  a re-examination of the reference cosmic
abundances and a detailed study of their local dependencies.

\subsection{Iron as a component of multi-layered particles}\label{fe}

Iron being highly depleted  in the interstellar medium
is one of the major dust-forming elements
(Snow \& Witt \cite{sw96}, Jones \cite{jones99}).
According to the recent results of Snow et al.~(\cite{srf02})
obtained with the Far Ultraviolet Spectroscopic Explorer
(FUSE), the abundance of iron in the dust-phase is in the range
from 95.2\% to 99.6\% of the cosmic abundance.
Iron can be incorporated into dust
grains in the form of oxides (FeO, Fe$_2$O$_3$, Fe$_3$O$_4$),
(Mg, Fe)-silicates, sulfides and metallic iron.
The last possibility causes from the theory of dust condensation
in  circumstellar environments (Lewis \&  Ney \cite{ln79},
Gail \&  Sedlmayr \cite{gs99}).
Gail \&  Sedlmayr (\cite{gs99}) note that  iron starts to condense
at a temperature well below the stability limits of some silicates
like forsterite and quartz. This leads to condensation of iron on the
surface of silicate grains where  iron islands can be formed.
The continuing growth of the islands will be accompanied by
condensation of new silicate layers and, finally, a layered particle
arise. The fraction of iron incorporated into grains
in the form of metallic Fe is inversely proportional to the stellar
mass-loss rate (Ferratori \& Gail \cite{fg01}).

Metallic iron was considered several times as a
separate component of dust mixtures
in the modelling of spectral energy distribution of OH/IR stars
(Harwit et al. \cite{hetal01}, Kemper et al. \cite{kkw02}),
Herbig Ae/Be stars  (Bouwman et al. \cite{bou00}),
protostellar objects (Demyk et al. \cite{dem99}).
The main reason of the inclusion of iron is its great opacity in the
near-infrared
spectral range. Usually the required mass fraction of metallic
iron is several percent (Bouwman et al. \cite{bou00}, Kemper et al. \cite{kkw02}).
However, sometimes iron is assumed to be the main component of the dust mixture
(Harwit et al. \cite{hetal01}).
In order to increase the opacity of iron particles further
and correspondingly
to decrease the mass fraction of iron the
shape of grains was considered to be needle-like (Kemper et al. \cite{kkw02})
that strongly increases the radiation pressure on them
and cuts down their lifetime in stellar atmospheres
(Il'in \& Voshchinnikov \cite{iv98}).
Needle-like iron dust has been also suggested to be present
in the intergalactic space and to termalize the cosmic
microwave background radiation, possibly arising from the
radiation of Population III stars (see Li \cite{li02} and references therein).
Note  that in many cases the optical properties of iron grains
were calculated in the Rayleigh approximation which however does not work
for iron in the near-infrared because of its high refractive index.
For example, at the wavelength $\lambda=5\,\mkm$,
$m_{\rm Fe}=4.59+15.4i$ (Lynch \& Hunter \cite{palik?})
and the Rayleigh criterion $|m|x \ll 1$ is satisfied only for
tiny particles ($r_{\rm s} \ll 0.05\,\mkm$).

\begin{figure}[htb]\bc
\resizebox{\hsize}{!}{\includegraphics{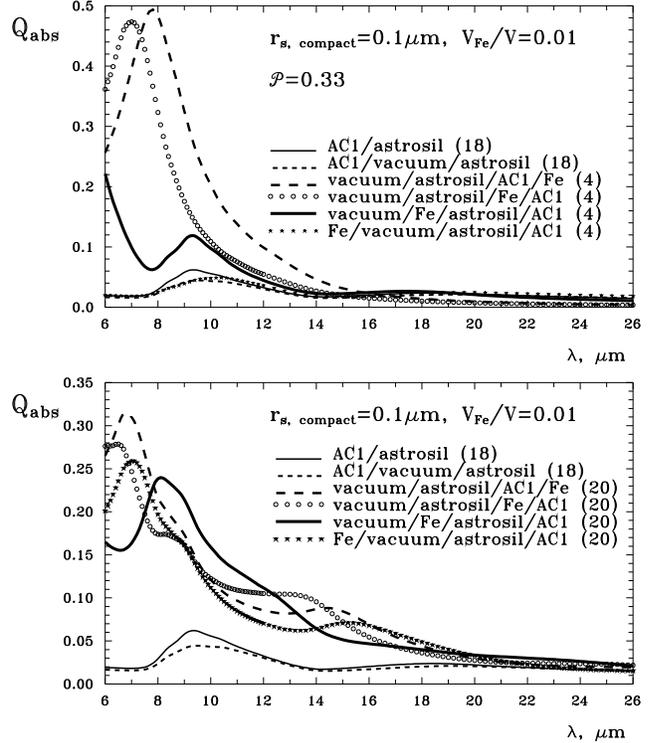}}
\caption{Wavelength dependence  of the absorption efficiency factors
for multi-layered spheres
of radius $r_{\rm s, \, compact}=0.1 \,\mkm$.
Compact particles contain an equal fraction (50\%)
of AC1 and astrosil.
Composite particles contain an equal fraction (33\%)
of AC1, astrosil and  vacuum, the volume fraction of iron is 1\%.
All constituents are separated in equivolume  layers.
The cyclic order of the layers and their number are indicated.
}\label{fe01}\ec\end{figure}

In order to consider the influence of the metallic iron on
the optical properties of multi-layered spheres we calculated
the wavelength dependence  of the absorption efficiency factors
similar to  what had been shown in Fig.~\ref{abs01}. Some results are
plotted in Fig.~\ref{fe01}. In all cases the volume fraction of iron
was adopted to be equal to 1\% of the total grain volume
(or approximately 4\% of grain mass).
The iron was put in the form of one layer at different
places inside the four-layered particle
(Fig.~\ref{fe01}, upper panel) or distributed as equivolume layers
inside the 20-layered particle (Fig.~\ref{fe01}, lower panel).
As it can be seen, the presence of metallic iron at any place
inside a particle, excluding its core, drastically changes
the particle absorption. Iron totally screens the underlying
layers and influences the optics of the overlying ones.
The silicate features disappear and a very strong ``pseudo-feature'' with
the centre between $6\,\mkm$ and $9\,\mkm$ arises instead of them.
Such a feature was not
observed in  spectra of celestial objects, and hence
 definitive doubts appear that  metallic iron is
a noticeable component of dust grains, at least as a layer.
However, iron can well form the core of a particle.
Very likely,  iron is oxidized in the result of redox
reactions or is modified to sulfides
(Duley \cite{dul80}, Jones \cite{jones90}).

\section{Concluding remarks}\label{concl}

We present a new model of composite grains which can be
used for the interpretation of observations of  interstellar,
circumstellar and cometary dust.
 In our model the particles are represented by multi-layered
spheres whose optical properties are calculated exactly.
 If the number of layers is small, the model obviously coincides
with older models where the grains have several coatings.
 For a large ($\ga 15-20$) number of the layers,
the new model is shown to approximate heterogeneous particles
consisting of several well mixed materials.

The model allows one a careful examination of the optical
properties of very porous particles
 Previously, this task was solved using the Mie theory
for homogeneous spheres and  effective refractive indices
derived from different mixing rules of the Effective Medium Theory.
 We demonstrate that this approach gives wrong results
when the porosity exceeds a value of 0.5.
 The application of a sophisticated layered-sphere mixing rule,
recently suggested by
Voshchinnikov \&  Mathis~(\cite{vm99}), provides results of acceptable accuracy.
 Our study of the optics of porous grains allows us to conclude that
a growth of porosity usually leads to:
\begin{itemize}
\item an increase of the extinction, scattering and absorption
      cross-sections (Figs.~\ref{fvac}, \ref{fvac2});
\item a growth of albedo (beginning with a size parameter value $2 - 3$)
      and the asymmetry parameter (for particles of all sizes,
      Fig.~\ref{fvac2});
\item flattening of the wavelength dependence of extinction (Fig.~\ref{ext_w});
\item an increase of the sweeping efficiency for small particles and
      its decrease for large particles (Fig.~\ref{betpr});
\item a growth of the intensity of radiation scattered in nearly forward
      directions and the transformation of the negative polarization
      to the positive one beginning with large scattering angles (Fig.~\ref{iipp});
\item a decrease of grain temperature (Fig.~\ref{td});
\item broadening of the infrared features and their shift to
      longer wavelengths (Fig.~\ref{abs01});
\item a growth of the particle opacity in the mm domain (Table~\ref{t1}).
\end{itemize}
Note that some of our results contradict the previous calculations
based on the approximate light scattering theory.

Application of our model to interpretation of the interstellar
extinction and cosmic abundances observed in the direction of two stars
shows that one can strongly reduce the model dust-phase abundances
and hence partly resolve the carbon crisis.
However, its final solution obviously needs
more work on the model as well as better known
reference cosmic abundances and their local variations.

We can also conclude that  metallic iron hardly is
a noticeable component of dust grains in the form of a layer
because its presence at any place inside a particle
drastically changes the particle absorption and leads
to nonobserved phenomena (Fig.~\ref{fe01}).
An exception is if iron is located in the core.

\acknowledgements{
We are thankful to Walter Wegner for the possibility to use unpublished data.
The work was partly supported by  grants
of the Volkswagen Foundation (Germany),  INTAS (Open Call 99/652),
and the DFG Research Group ``Laboratory Astrophysics''.
}


\end{document}